\documentclass[twocolumn]{aastex62}
\usepackage[utf8]{inputenc}
\DeclareUnicodeCharacter{2212}{-}
\usepackage{graphicx}
\graphicspath{{./}{Figures}}
\usepackage{amsmath}
\usepackage{ulem}
\usepackage{breqn}
\usepackage{afterpage}
\usepackage{enumitem}
\usepackage{xcolor}
\usepackage{float}
\usepackage{xspace}
\usepackage{listings}
\usepackage{url}
\usepackage{fontawesome}
\setenumerate{itemsep=0mm}
\usepackage{lineno}
\linenumbers
\usepackage{wrapfig}
\usepackage{placeins}
\usepackage{CJK}
\usepackage[hang,flushmargin]{footmisc}

\newcommand{\vv}{{\tablenotemark{\footnotesize{a}}}}
\newcommand{\xx}{{\tablenotemark{\footnotesize{b}}}}
\newcommand{\yy}{{\tablenotemark{\footnotesize{c}}}}
\newcommand{\eps}[1]{\ensuremath{\log\epsilon\,(\mathrm{#1})}}
\newcommand{\abund}[2]{\ensuremath{[\mathrm{#1}/\mathrm{#2}]}}
\newcommand{\xfe}[1]{\abund{#1}{Fe}}

\newcommand{\msun}{\ensuremath{M_\odot}\xspace}

\newcommand{\z}{\ensuremath{z}\xspace}

\newcommand{\teff}{\ensuremath{{T_{\rm eff}}}\xspace}
\newcommand{\logg}{\ensuremath{{\log g}}\xspace}
\newcommand{\vlos}{\ensuremath{{v_{\rm los}}}\xspace}
\newcommand{\vturb}{\ensuremath{{\xi}}\xspace}
\newcommand{\cfe}{\ensuremath{{\rm[C/Fe]}}\xspace}

\newcommand{\kmsec}{\ensuremath{{\rm km\,s^{-1}}}\xspace}
\newcommand{\angmom}{\ensuremath{{\rm kpc\,km\,s^{-1}}}\xspace}

\newcommand{\angs}{{\AA}\xspace}
\newcommand{\gdr}{GDR3\_479591\xspace} 
\newcommand{\mw}{Milky~Way\xspace} 
 
\definecolor{codegreen}{rgb}{0,0.6,0}
\definecolor{codegray}{rgb}{0.5,0.5,0.5}
\definecolor{codepurple}{rgb}{0.58,0,0.82}
\definecolor{backcolour}{rgb}{0.95,0.95,0.92}
\definecolor{backcolour}{rgb}{1.0,1.0,1.0}

\lstdefinestyle{mystyle}{
    backgroundcolor=\color{backcolour},   
    commentstyle=\color{codegreen},
    keywordstyle=\color{magenta},
    numberstyle=\tiny\color{codegray},
    stringstyle=\color{codepurple},
    basicstyle=\ttfamily\footnotesize,
    breakatwhitespace=false,         
    breaklines=true,                 
    captionpos=b,                    
    keepspaces=true,                 
    numbers=left,                    
    numbersep=5pt,                  
    showspaces=false,                
    showstringspaces=false,
    showtabs=false,                  
    tabsize=2
}

\usepackage{amsmath}
\usepackage{amssymb}
\usepackage{xspace}
\usepackage{xifthen}
\usepackage{eso-pic}

\definecolor{forestgreen}{HTML}{228B22}
\definecolor{urlblue}{HTML}{000000}

\mathchardef\mhyphen="2D

\newlength{\dhatheight}

\newcommand{\unit}[1]{\ensuremath{\mathrm{\,#1}}\xspace}

\newcommand{\e}{\unit{e^{-}}}

\newcommand{\bandvar}[2][]{%
  \ifthenelse{\isempty{#1}}{\var{#2}}{\var{#2\_#1}}%
}

\newcommand{\feh}{{\ensuremath{\rm [Fe/H]}}\xspace}

\newcommand{\var}[1]{\ensuremath{\texttt{\MakeUppercase{#1}}}\xspace}

\providecommand\physrep{\ref@jnl{Phys.~Rep.}}%
\providecommand\apjs{\ref@jnl{ApJS}}%
\providecommand{\jcap}{\ref@jnl{JCAP}}%

\shorttitle{A CNO-Enhanced UMP Star}
\shortauthors{Limberg et al.}

\begin{document}

\title{\textbf{
Chemo-dynamical Analysis of a CNO-Enhanced Ultra Metal-poor Star ($\feh < -4$): \\ Insights into Early Enrichment by Faint Population~III Supernovae\footnote{Based on observations gathered with the 6.5\,m Magellan \\ Telescopes located at Las Campanas Observatory, Chile.}
}}


%

\newcommand{\MITall}{Department of Physics and Kavli Institute for Astrophysics and Space Research, Massachusetts Institute of Technology, 77 Massachusetts Avenue, Cambridge, MA 02139, USA}
\newcommand{\MITPhysics}{Department of Physics, Massachusetts Institute of Technology, 77 Massachusetts Avenue, Cambridge, MA 02139, USA}
\newcommand{\MITKavli}{Kavli Institute for Astrophysics and Space Research, Massachusetts Institute of Technology, 77 Massachusetts Avenue, Cambridge, MA 02139, USA}
\newcommand{\UChicagoAA}{Department of Astronomy \& Astrophysics, University of Chicago, 5640 S Ellis Avenue, Chicago, IL 60637, USA}
\newcommand{\KICP}{Kavli Institute for Cosmological Physics, University of Chicago, 5640 S Ellis Avenue, Chicago, IL 60637, USA}
\newcommand{\IAGUSP}{Universidade de S\~ao Paulo, Instituto de Astronomia, Geof\'isica e Ci\^encias Atmosf\'ericas, Departamento de Astronomia, SP 05508-090, S\~ao Paulo, Brazil}
\newcommand{\NOIRLab}{NSF NOIRLab, Tucson, AZ 85719, USA}
\newcommand{\JINA}{Joint Institute for Nuclear Astrophysics--Center for the Evolution of the Elements (JINA-CEE), USA}

\newcommand{\stanfordKITP}{Kavli Institute for Particle Astrophysics \& Cosmology, P.O. Box 2450, Stanford University, Stanford, CA 94305, USA}


\correspondingauthor{Guilherme Limberg}
\author[0000-0002-9269-8287]{Guilherme~Limberg}
\email{limberg@uchicago.edu}
\affiliation{\KICP}
\affiliation{\UChicagoAA}

\author[0000-0003-4479-1265]{Vinicius M.\ Placco}
\email{}
\affiliation{\NOIRLab}

\author[0000-0002-4863-8842]{Alexander~P.~Ji}
\email{}
\affiliation{\UChicagoAA}
\affiliation{\KICP}

\author{Yupeng Yao}
\email{}
\affiliation{Department of Computer Science and Engineering, University of North Texas, 3940 N. Elm Street, Denton, TX 76207, USA}

\author[0000-0003-4524-9363]{Friedrich Anders}
\email{}
\affiliation{Departament de Física Quàntica i Astrofísica
, Universitat de Barcelona
, C Martí i Franquès, 1, 08028 Barcelona, Spain}
\affiliation{Institut de Ci\`encies del Cosmos, Universitat de Barcelona
, C Mart\'i i Franqu\`es 1, 08028 Barcelona, Spain}
\affiliation{Institut d’Estudis Espacials de Catalunya
, Edifici RDIT, Campus UPC, 08860 Castelldefels (Barcelona), Spain}

\author[0009-0007-3791-7890]{Wendy Q. Sun}
\affiliation{\MITall}
\email{}

\author[0000-0002-2380-9801]{Anna de Graaff}
\thanks{Clay Fellow}
\affiliation{Center for Astrophysics $|$ Harvard \& Smithsonian, 60 Garden St., Cambridge MA 02138 USA}
\affiliation{Max-Planck-Institut f\"ur Astronomie, K\"onigstuhl 17, D-69117 Heidelberg, Germany}
\email{}

\author[0000-0003-3997-5705]{Rohan P. Naidu}
\affiliation{\MITall}
\email{}

\author[0000-0003-1561-3814]{Harley Katz}
\email{}
\affiliation{\UChicagoAA}
\affiliation{\KICP}

\author[0000-0002-3867-3927]{Pierre~N.~Thibodeaux}
\email{}
\affiliation{\UChicagoAA}
\affiliation{\KICP}

\author[0000-0002-7155-679X]{Anirudh Chiti}
\email{}
\affiliation{\stanfordKITP}

\author[0000-0001-9178-3992]{Mohammad K.\ Mardini}
\email{}
\affiliation{Department of Physics, Zarqa University, Zarqa 13110, Jordan}
\affiliation{\JINA}

\author[0000-0002-2139-7145]{Anna Frebel}
\email{}
\affiliation{\MITall}
\affiliation{\JINA}

\begin{abstract}

We present a comprehensive chemo-dynamical analysis of the previously known ultra metal-poor (UMP) star ($\feh =-4.06$) \textit{Gaia} DR3 \texttt{source\_id} 4795913112968206720 (\gdr), but independently recovered from the \textit{Gaia} mission's Blue and Red Photometer `XP' spectro-photometric catalog. We combine multi-band photometry, astrometry, and high-resolution spectroscopy to confirm \gdr as a red giant-branch (RGB) star 
located at $\sim$6\,kpc 
from the Sun. Abundance analysis under local thermodynamic equilibrium reveals significant enhancements, relative to the solar level, in carbon ($\cfe = +1.54$, after evolutionary correction for depletion in the RGB), nitrogen ($\rm[N/Fe] = +2.64$), and oxygen ($\rm[O/Fe] = +2.62$). The CNO-enhanced \gdr is thus one of only 5 UMP stars with a detected oxygen abundance. The CNO excess is accompanied by enhancements in several other light elements, such as Na, Mg, Al, and Si. We demonstrate that the chemical pattern of \gdr can be reproduced by the yields of a single Population~III `faint' supernova with progenitor mass of $\sim$21-to-28\,\msun and a low explosion energy of ($0.3 \leq E_{\rm SN}/(10^{51}\,{\rm erg}) \leq 0.9$). Additionally, we use literature metal-poor stars to show that, contrary to recent propositions for high-redshift galaxies, a mild enhancement in [C/O] does not automatically translate to the high [C/Fe] typically observed in UMP stars in the Milky Way and its satellites. \gdr could not be dynamically associated with any of the most relevant accreted substructures in the Galactic halo, and we speculate that it was formed in an ultra-faint dwarf galaxy environment that later merged with the Milky Way.

\end{abstract}


\section{Introduction} \label{sec:intro}

\setcounter{footnote}{0}

To constrain the properties of the first Population~III stars is the overarching goal of Stellar Archaeology \citep[][]{frebel2015}. These first metal-free stars illuminated the first galaxies at high redshift ($z \gtrsim 15$) with ionizing photons \citep[][]{BrommYoshida2011review}, bring the Universe out of the cosmic dark ages and into the era of reionization \citep[][]{BarkanaLoeb2001reionization, robertson2022reionization}. At the same time, the radiative feedback produced by massive Population~III stars is expected to regulate subsequent star-formation in low-mass halos via, for example, dissociation of molecular hydrogen (${\rm H}_2$), the primary coolant in metal-poor gas, potentially even shaping the luminosity function of relic dwarf galaxies observed in the Local Group \citep[see the review by][as well as \citealt{brauer2025aeos}]{klessen2023pop3Rev}. Additionally, the associated first supernovae contributed to the earliest chemical enrichment episodes \citep[][]{HegerWoosley2002} whose metals were likely responsible for the transition from a primordial top-heavy initial mass function to a bottom-heavy one that we see today \citep[][]{omukai2005imf, marks2012imf_metallicity, chon2021MetalPoorIMF}.

The exciting possibility of directly observing metal-free material in high-redshift galaxies has been reinvigorated by the launch of \textit{JWST}, especially when aided by gravitational lensing magnification \citep[e.g.,][]{vanzella2023lap1b}. Having said that, the most chemically pristine systems identified in this way so far are primeval dwarf galaxies (${\rm stellar \ masses } < 10^6\,\msun$) with metallicity values that barely reach 0.1\% of the solar level \citep[][]{fujimoto2025pop3Gal, Morishita2025pop3Gal}. Perhaps this limitation is not unexpected given that metal-enriched galaxies at redshift $z\geq 10$ should exhibit weak [\ion{O}{3}] 5007\,\angs emission lines that are difficult to detect even at \textit{JWST}'s sensitivity \citep[][]{katz2023pop3}, i.e., masquerading these systems as Population~III environments. On the other hand, the Stellar Archaeology approach has provided discoveries of ultra metal-poor stars (UMP, $\feh <-4$\footnote{Definition of elemental abundance for a given star ($\star$) relative to the Sun ($\odot$): ${\rm[X/Y]} = \log{(N_{\rm X}/N_{\rm Y})_\star} - \log{(N_{\rm X}/N_{\rm Y})_\odot}$, where $N_{\rm X}$ ($N_{\rm Y}$) is the number density of atoms of element X (Y).
\label{footnote_abund}}) well below 0.01\% solar mass-fraction metallicity. The most notable examples are stars SDSS~J102915$+$172927 \citep{caffau2011} and GDR3\_526285 \citep[][]{limberg2025ump_gaiaxp, ji2026natas}, both confidently at $Z/Z_\odot < 0.01\%$.

To connect local UMP stars to their pristine Population~III progenitors, the established go-to method is to compare their detailed elemental abundance patterns with the predicted yields for metal-free supernova nucleosynthesis models \citep[][]{Lai2008_sn_yield_fits, placco2015seguefollowup, Placco2016B}. The key inferred properties from this approach are the progenitor masses and associated supernova explosion energies \citep[e.g.,][]{Ishigaki2018, Jiang2025abundFitbayesian}. One obvious limitation on the theoretical side is that currently available yield models certainly do not encompass the full range of possible parameters regarding explosion physics, nucleosynthesis networks, and stellar evolution prescriptions \citep[][]{AlexJi2024spectacular}. On the observational front, the main caveat of the Stellar Archaeology concept is that, since surviving UMP stars are extremely rare, their identification is difficult with only $\sim$40 such objects known \citep[][]{sestito2019, placco2021ump}, hence averaging $\sim$1 discovery per year since \citet[][]{Bessel1984}. Moreover, even with high-resolution ($\mathcal{R} \geq 20k$) stellar spectroscopy, not all elements useful for direct comparisons with high-redshift galaxies are typically detected. For example, only 5 out of the known UMP stars have a measured oxygen abundance according to our own curation (Section \ref{subsec:lit_ump})
.


Out of this limited sample, the most obviously notable fact is that no surviving Population~III star has ever been discovered. Having said that, \citet[][]{hartwig2015imf} estimated that, to rule out the existence of low-mass ($\leq$0.8\,\msun) metal-free stars in the Milky Way at 95\% confidence, we would need to observe $\sim$400 UMP stars, although stringent constraints might also come from lower-mass environments such as dwarf satellite galaxies \citep[][]{magg2018, rossi2021pop3imfUFD}. In this context, the most exceptional feature of UMP stars is that $\gtrsim$80\% of them are over-enriched in carbon by a factor of $>$5 compared to the solar level \citep[][]{placco2014Carbon, Arentsen2022cemp}, i.e., they belong to the category of carbon-enhanced metal-poor stars \citep[CEMP;][]{beers2005, aoki2007}. This CEMP signature has been interpreted as a key nucleosynthesis byproduct of the first stars \citep[][]{cooke2014cemp, vanni2023_SNenergy}, particularly through low-energy explosions in `faint' supernovae (see \citealt{Nomoto2013} for a review). The discovery of a CEMP UMP star in a $\sim$2000\,$M_\odot$ dwarf galaxy has provided strong corroboration to this hypothesis \citep[][]{chiti2026pic2}. Also, a CEMP environment has even been tentatively identified at redshift $z\sim6.6$ with \textit{JWST} \citep[][]{nakajima2025pop3Gal}, presumably connecting to local CEMP stars \citep[][]{deugenio2024gsz12, pollock2026highz}.

The most successful modern surveys\footnote{Other relevant projects with low-resolution spectroscopy in the past two decades include the Hamburg/ESO survey \citep[][]{Wisotzki1996hamburgESO, Christlieb2008hes}, the Sloan Digital Sky Survey \citep[][]{sdssYork}, and the Large Sky Area Multi-object Fiber Spectroscopic Telescope \citep[][]{LAMOST1}.} for UMP stars rely on narrow-band photometric data around the \ion{Ca}{2} K and H lines at 3934\,\angs and 3969\,\AA, respectively, which is reminiscent of earlier efforts \citep[][]{bond1970, bidelman1973, beers1985}. We highlight the \textit{Pristine} survey conducted from the northern hemisphere \citep[][]{Starkenburg2017PRISTINE} and, from the south, SkyMapper \citep[][]{keller2007skymapper}, the Southern Photometric Local Universe Survey \citep[SPLUS;][including its short-exposure component by \citealt{splusSHORTSdr1}]{splus}, as well as the newest Mapping the Ancient Galaxy in CaHK \citep[MAGIC;][]{Chiti2026magicOverview}. Indeed, each of these programs have contributed their own discoveries to our current census of UMP stars \citep[][]{keller2014, starkenburg2018ump, nordlander2019smss, placco2021ump, placco2025magic, chiti2026pic2}.

With the advent of the \textit{Gaia} space mission \citep{GaiaMission}, we are now capable of extending the hunt for UMP stars to the whole sky with homogeneous data. In particular, with the mission's third data release \citep[DR3;][]{GaiaDR3} and publication of blue (\textit{BP}) and red (\textit{RP}) photometer low-resolution `XP' spectra \citep[$\mathcal{R} \sim 50$;][]{DeAngeli2023gaiaxp}, we can perform narrow-band \textit{Pristine}/SkyMapper/SPLUS/MAGIC-like searches for UMP stars and, at the same time, contextualize new findings with \textit{Gaia}-based kinematics \citep[e.g.,][]{Yuan2020dtgs, Limberg2021dtgs, Shank2022dtgs}. The usability of \textit{Gaia}~XP spectra for the task has been explored in some works \citep[][]{Viswanathan2025pristineGaia, YeXianhao2025gaiaxp, thai2026gaiaxp}, including 
spectroscopic validation. So far, the most outstanding discoveries 
include the lowest-metallicity star in the Large Magellanic Cloud \citep[LMC-119;][]{chiti2024lmc}, and the aforementioned GDR3\_526285 UMP object described in \citet[][also \citealt{ji2026natas}]{limberg2025ump_gaiaxp}.

In this contribution, we report the independent recovery, spectroscopy, and chemo-dynamical analysis of \textit{Gaia}~DR3 \texttt{source\_id} 4795913112968206720 (hereafter ``GDR3\_479591''), a chemically peculiar red giant-branch (RGB) UMP star ($\feh = -4.06$) in \textit{Gaia}~XP. We note that \citet[][]{DaCosta2019} had previously correctly identified \gdr as an RGB UMP candidate, then vetted by \citet[][]{yong2021_SMSSmetalpoor}\footnote{Named SMSS~J054650.97-471407.9 following the SkyMapper survey nomenclature \citep{keller2007skymapper}.} with high-resolution spectroscopy. In this work, we detect additional elements, derive an independent distance estimate, perform a dynamical investigation, fit Population~III supernova yields to the complete abundance pattern of \gdr, and contextualize our findings with Milky Way halo stellar populations as well as recent discoveries of high-redshift galaxies with \textit{JWST}. We describe the astrometric, photometric, and spectroscopic data in Section \ref{sec:data}. Our methods are shown in Section \ref{sec:methods}. Results and conclusions are listed in Sections \ref{results} and \ref{conclusions}, respectively.

\renewcommand{\arraystretch}{1.0}
\setlength{\tabcolsep}{0.1em}
\begin{table*}[ht!]
\centering
\caption{Observational Data for \textit{Gaia} DR3 4795913112968206720
}
\label{tabelao}
\begin{tabular}{>{\normalsize}l >{\normalsize}c >{\normalsize}r >{\normalsize}l >{\normalsize}l}
\hline
\hline
Quantity & Symbol & Value & Unit & Source and/or reference\\ %
\hline
\hline
Right ascension & $\alpha$ & 86.7124 & degree & \textit{Gaia} DR3 \\ 
Declination & $\delta$ & $-$47.2355 & degree & \textit{Gaia} DR3 \\ 
Galactic longitude & $\ell$ & 253.9486 & degree & \textit{Gaia} DR3 \\ 
Galactic latitude & $b$ & $-$30.1865 & degree & \textit{Gaia} DR3 \\ 
Parallax & $\varpi$ & $0.135\pm 0.012$ & mas & \textit{Gaia} DR3, \citet[][]{ElBadry2025plx} \\ 
Parallax zero point correction & $\varpi_{\rm zpt}$ & $-$0.033 & mas & \citet[][]{Lindegren2021_PlxBias} \\ 
Proper motion ($\alpha$) & $\mu_{\alpha,*}$\footnote{$\mu_{\alpha,*} = \mu_{\alpha} \cos \delta$} & $-4.186\pm0.012$ & mas yr$^{-1}$ & \textit{Gaia} DR3 \\ 
Proper motion ($\delta$) & $\mu_\delta$ & $-2.720\pm0.013$ & mas yr$^{-1}$ & \textit{Gaia} DR3 \\ 
Re-normalized unit weight error & \texttt{ruwe} & 1.032 &  & \textit{Gaia} DR3 \\ 

\hline 
Mass & $M$ & $0.75\pm0.15$ & $M_\odot$ & Assumed (this work) \\ 

\hline
$B$ magnitude & $B$ & 14.599 & & \textit{Gaia} XP spectra synthetic photometry \\ 
$V$ magnitude & $V$ & 13.911 & & \textit{Gaia} XP spectra synthetic photometry \\

$G$ magnitude & $G$ & 13.663 & & \textit{Gaia} DR3 \\ 
$BP$ magnitude & $BP$ & 14.082 & & \textit{Gaia} DR3 \\ 
$RP$ magnitude & $RP$ & 13.065 & & \textit{Gaia} DR3 \\ 

$J$ magnitude & $J$ & 12.244 &  & 2MASS \\
$H$ magnitude & $H$ & 11.853 & & 2MASS \\
$K$ magnitude & $K_s$ & 11.713 &  & 2MASS \\



\hline 
Color excess & $E(B-V)$ & $0.052 \pm 0.002$ & & \citet[][]{Schlafly2011} \\ 
Extinction ($V$) & $A_V$ & $0.160 \pm 0.005$ & & $A_V = R_V \cdot E(B-V)$, with $R_V = 3.1$ \\ 
Extinction (other filters, $\lambda$) & $A_\lambda$ & Many values & & \citet{WangChen2019extinctions} \\ 
Bolometric correction ($G$) & ${\rm BC}(G)$ & $-0.181 \pm 0.003$ & & \citet[][]{CasagrandeVandenBerg2018gaia} \\

\hline 
\textbf{Line-of-sight velocity} & $v_{\rm los}$ & ${+}267.5 \pm 1.5$ & km\,s$^{-1}$ & \textbf{This work} \\
 &  & ${+}267.6 \pm 4.6$ & km\,s$^{-1}$ & \textit{Gaia} DR3 \\
&  & ${+}267.2$ & km\,s$^{-1}$ & \citet[][]{yong2021_SMSSmetalpoor} \\

\hline 
\textbf{Effective temperature} & $\teff$ & $5181 \pm 31$ & K & \textbf{This work} \\
&  & 5175 & K & \citet[][]{yong2021_SMSSmetalpoor} \\
&  & 5116 & K & \citet[][]{andrae2023gaiaXP} \\

\textbf{Log of surface gravity} & $\logg$ & $2.02 \pm 0.11$ & [cgs] & \textbf{This work} \\
&  & 2.40 & [cgs] & \citet[][]{yong2021_SMSSmetalpoor} \\
&  & 2.23 & [cgs] & \citet[][]{andrae2023gaiaXP} \\

\textbf{Microturbulence velocity} & \vturb & $1.55 \pm 0.15$ & km\,s$^{-1}$ & \textbf{This work} \\
& & 1.8 & km\,s$^{-1}$ & \citet{yong2021_SMSSmetalpoor} \\

\textbf{Metallicity} (iron abundance) & \feh & ${-}4.06 \pm 0.17$ & & \textbf{This work} \\
&  & $-$3.4 & & \citet[][]{yao2024gaiaxp}, \citet[][]{limberg2025ump_gaiaxp} \\
&  & $-$4.09 & & \citet[][]{yong2021_SMSSmetalpoor} \\
&  & $-$3.2 & & \citet[][]{andrae2023gaiaXP} \\

\hline 
Distance modulus & $\mu$ & $13.87^{+0.16}_{-0.15}$ &  & This work \\
\textbf{Heliocentric distance} & $d_{\rm h}$ & $5.95^{+0.45}_{-0.39}$ & kpc & \textbf{This work} \\ 
&  & $5.86^{+0.43}_{-0.30}$ & kpc & \citet[][]{BailerJones2021gaiaEDR3}, geometric \\
&  & $5.64^{+0.42}_{-0.49}$ & kpc & \citet[][]{Anders2022starhorseEDR3}, \texttt{StarHorse} \\
\hline
Orbital energy & $E$ & $(-1.34 \pm 0.03)\times10^5$ & ${\rm km}^2\,{\rm s}^{-2}$ & This work \\
Apocenter & $r_{\rm apo}$ & $13.9^{+1.2}_{-1.4}$ & kpc & This work \\
Pericenter & $r_{\rm peri}$ & $11.1\pm0.3$ & kpc & This work \\
Eccentricity & $e$ & $0.11\pm0.03$ & & This work \\
Inclination & $\theta$ & $101.5\pm0.6$ & deg & This work \\
$x$ component of angular momentum & $L_x$ & $1380^{+127}_{-121}$ & kpc\,km\,s$^{-1}$ & This work \\
$y$ component of angular momentum & $L_y$ & $-2351^{+100}_{-105}$ & kpc\,km\,s$^{-1}$ & This work \\
$z$ component of angular momentum & $L_z$ & $-555\pm8$ & kpc\,km\,s$^{-1}$ & This work \\
\hline
\end{tabular}
\end{table*}

\section{Data} \label{sec:data}

\subsection{Target selection}

\gdr was originally selected as a very metal-poor RGB star ($\feh \leq -2$) by \citet[][]{yao2024gaiaxp}. These authors developed a machine-learning classification algorithm to consolidate their best sample of $>$70,000 very metal-poor candidates from \textit{Gaia} XP data. Then, we derive data-driven \feh estimates for the \citet[][]{yao2024gaiaxp} sample with a regression model trained on high-resolution ($\mathcal{R}\geq 20k$) spectroscopic catalogs of metal-poor stars in the literature \citep[][]{Cayrel2004, Cohen2013, roederer2014, Jacobson2015metalpoor, holmbeck2020, Li2022lamost}, as described in \citet[][]{limberg2025ump_gaiaxp}. We find $\feh = -3.4$ for \gdr, which, in combination with the value of $\feh = -3.2$ provided by \citet[][also with \textit{Gaia} XP]{andrae2023gaiaXP}, led this star to be a high-priority target for spectroscopic follow-up.

Additional selection criteria include \gdr's color consistent with an RGB status\footnote{The color cut adopted by \citet{yao2024gaiaxp} is ${BP}_0-{RP}_0 > 0.8$.} \citep[${BP}_0-{RP}_0 = 0.95$ in \textit{Gaia} photometric system;][]{Jordi2010_GaiaPhotoSystem} after reddening/extinction correction, as indicated by the `0' subscripts (see below), and relatively high brightness in the blue portion of its spectral energy distribution (${BP} = 14.1$), where many absorption features of interest to Stellar Archaeology are located. Estimated stellar parameters from \textit{Gaia} XP by \citet[][]{andrae2023gaiaXP} are also consistent with an RGB solution, namely effective temperature $\teff = 5116\,{\rm K}$ and logarithm of surface gravity $\logg = 2.23$. Lastly, the previous high-resolution spectroscopic analysis by \citet{yong2021_SMSSmetalpoor} confirms that \gdr is a UMP RGB star ($\teff = 5175\,{\rm K}$, $\logg = 2.40$, $\feh = -4.09$).

\subsection{Available photometry and astrometry} \label{photometry_astrometry}

We gather multi-band optical and infrared photometric data for \gdr from both \textit{Gaia} DR3 \citep[$G$, ${BP}$, and ${RP}$;][]{GaiaEDR3Photometry} and the Two Micron All Sky Survey \citep[2MASS; $J$, $H$, and $K_s$;][]{2MASS}. We also generate synthetic $B$ and $V$ photometry in the standardized Johnson-Kron-Cousins (JKC) system \citep[e.g.,][]{Landolt1992ubvri, BessellMurphy2012ubvri} for \gdr from its \textit{Gaia} XP spectra using the \texttt{GaiaXPy} toolkit\footnote{\url{https://gaiaxpy.readthedocs.io/}.} \citep[][]{Montegriffo2023_GaiaXPy}. 

Although \gdr is located far from the Galactic plane (Galactic latitude $b \sim -30\,{\rm deg}$; see Table \ref{tabelao}), we correct all photometric data for the effects of interstellar dust;
\begin{equation}
    m_{\lambda, 0} = m_\lambda - A_\lambda,
\end{equation}
where $m_{\lambda, 0}$ is the dereddened $m_\lambda$ apparent magnitude at filter $\lambda$ and $A_\lambda$ is the extinction at the same band. We adopt the reddening $E(B-V) = 0.052\pm0.002$ value from \citet[][]{Schlafly2011} with the \citet[][]{Schlegel1998} dust map. The linear extinction law is defined as 
\begin{equation}
    A_\lambda = R_\lambda\cdot E(B-V)
\end{equation}
and the $V$-band extinction-to-reddening ratio is assumed to be $R_V = 3.1$. For all other magnitudes, we take $A_\lambda/A_V$ values, hence $R_\lambda$, from \citet[][]{WangChen2019extinctions}.

On the astrometry side, we recover proper motions, parallax ($\varpi$), and associated Gaussian uncertainties from \textit{Gaia} DR3. We compute the parallax zero-point bias ($\varpi_{\rm zpt}$) for \gdr using the formula provided in \citet[][]{Lindegren2021_PlxBias}\footnote{\url{https://pypi.org/project/gaiadr3-zeropoint/}.}. The final corrected parallax is given by 
\begin{equation}
    \varpi_{\rm cor} = \varpi - \varpi_{\rm zpt},
\end{equation}
which then serves as input for distance determinations below. We also employ the empirical prescription by \citet[][]{ElBadry2025plx} to inflate the parallax uncertainty as a function of the re-normalized unit weight error quantity (\texttt{ruwe}) provided in the \textit{Gaia} DR3 catalog. All observational information is organized in Table \ref{tabelao}. 

\subsection{Observations, data reduction, line-of-sight velocity determination, and spectrum normalization} \label{obs}

We obtained a high-resolution spectrum of \gdr with the Magellan Inamori Kyocera Echelle (MIKE; \citealt{Bernstein2003mike}) instrument at the Magellan Clay 6.5\,m telescope located in Las Campanas, Chile, on March 14th 2026 under good conditions (${\rm seeing} \sim 0.8\,{\rm arcsec}$). The observing setup included 2$\times$2 on-chip binning and a 0.7\,arcsec slit, yielding resolving powers of ${\sim}35k$/$28k$ in the blue/red arms of MIKE spectrum and a combined coverage of ${\sim}[3300{:}9000]$\,\AA. The single 20\,min exposure with this configuration during twilight reached signal-to-noise ratios of $S/N = 36$, 45, and 75 at 4000\,\AA, 5000\,\AA, and 6500\,\AA, respectively.

We perform data reduction with standard routines developed for MIKE with the \texttt{CarPy} package \citep[][]{carpy}, consisting of bias subtraction, quartz and diffuse `milky' flats, and wavelength calibration with a ThAr arc lamp. We compute the line-of-sight velocity (\vlos) by cross-correlating against a high-$S/N$ MIKE spectrum of metal-poor standard HD~122563 \citep[$\feh = -2.75$;][]{Karovicova2020standards} with the Spectroscopy Made Harder software \citep[\texttt{smhr};][]{smhr}. We derive \vlos in three different spectral windows; within [8450:8750]\,\angs covering the near-infrared \ion{Ca}{2} triplet, at [5100:5200]\,\angs for the \ion{Mg}{1}\,b triplet, and at [6510:6610]\,\angs with H$\alpha$. We take the average between these independent estimates and add the barycentric correction to obtain the final $\vlos = 267.5\pm1.5\,\kmsec$. The uncertainty is the quadrature sum of the standard deviation of individual measurements (1.1\,\kmsec) plus MIKE's systematic error floor due to slit centering and wavelength calibration (1.0\,\kmsec). Our determination is essentially identical, though more precise, to the reported \textit{Gaia}~DR3 value of $267.6\pm4.6\,\kmsec$ (Table \ref{tabelao}) as well as consistent with $267.2\,\kmsec$ from \citet[][]{yong2021_SMSSmetalpoor}\footnote{These authors did not report individual velocity uncertainties in their study, but informed an average 0.5\,\kmsec statistical error in their sample of 150 metal-poor stars, without information about systematics. Given that \gdr has the lowest metallicity out of those, this value should be taken as a lower limit.}. With the reduced, velocity-corrected data at hand, we finally stitch orders with variance weighting and normalize the MIKE spectrum of \gdr using cubic spline functions in the \texttt{smhr} environment.

\subsection{A new curated list of literature UMP stars} \label{subsec:lit_ump}

For comparison with \gdr, we curate a comprehensive list of UMP stars from the literature. We initiate by collecting data for $\feh <-3$ stars from both \texttt{SAGA} \citep[][]{suda2008} and the updated \texttt{JINAbase}\footnote{\url{https://github.com/Mohammad-Mardini/JINAbase-updated}.} \citep{jina} with measurements from high-resolution spectroscopic studies ($\mathcal{R}\geq 15k$). We then search for additional UMP stars using the NASA Astrophysics Data System service\footnote{\url{https://ui.adsabs.harvard.edu/}.}. This procedure allows us to identify some UMP stars that appear neither in \texttt{SAGA} nor \texttt{JINAbase} \citep[e.g.,][]{GonzalezHernandez2020}. Then, we compare our list with the \citet{sestito2019} compilation to ensure we have recovered all UMP stars from that work. In the final list, we keep only 50 (43) stars with reported $\feh < -3.95$ ($-4.00$), including \gdr itself (this work). We keep stars with up to $+0.05$\,dex above the formal definition in [Fe/H] simply to encapsulate plausible UMP objects given typical uncertainties. We caution the reader that these exact numbers could change if non-local thermodynamic equilibrium (NLTE) effects were accounted for \citep[][]{Ezzeddine2017nlte, NorrisYong2019nlte}, but, in our compilation, we only report LTE values. Also, we reject candidate UMP stars from \citet[][]{aguado2016sdssFollowup, aguado2017gtc, aguado2017wht} that have only moderate-resolution spectroscopy available, but were present in the \citet[][]{sestito2019} list. Several UMP stars have been studied by different authors, and we inherit the priority schemes implemented in \texttt{SAGA} and \texttt{JINAbase} for the adopted references in Appendix \ref{appendix:ump_stars} (Table \ref{tab:ump_comp}). 

Here, we provide notes on individual stars. Object HE~0057$-$5959, listed $\feh = -4.08$ in \citet[][ following \citealt{norris2013}]{sestito2019}, has a higher iron-abundance estimate in \citet[][$\feh = -3.94$]{Jacobson2015metalpoor} and is, therefore, removed. Star BD$+$44~493 was not originally present in our \texttt{SAGA} and \texttt{JINAbase} retrievals given its $\feh = -3.83$ value in \citet[][]{Ito2013}. A more recent analysis, however, has BD$+$44~493 at $\feh = -3.96$ \citep[][]{placco2024}, hence it is included in Table \ref{tab:ump_comp}. The opposite happens to star J1253$+$0753, which had $\feh = -4.02$ estimated by \citet[][]{LiHaining2015lamost}, but now \citet[][]{Li2022lamost} has it at $\feh = -3.83$. We also notice that \citet[][]{cohen2008} has a discrepant $\feh = -3.43$ value for HE~1012$-$1540, but we retain it in Table \ref{tab:ump_comp} given \citet[][$\feh = -4.17$]{roederer2014} parameters. Star J2050$-$6613 has a reported $\feh = -3.85$ value in \citet[][]{yong2021_SMSSmetalpoor}, but a more recent estimate by \citet[][]{mardini2024diskUMP} has it at $\feh = -4.05$, so we keep it. Ignoring upper limits, the CEMP fraction ($\rm[C/Fe] > +0.7$) in our curated sample of UMP stars is $86^{+8}_{-13}\%$. If we consider all UMP stars with only \cfe upper limits available to be non-CEMP, we find $70^{+12}_{-15}\%$. We caution that this latter choice removes some UMP stars from the CEMP category with quite loose carbon-abundance upper limits such as HE~2239$-$5019 at $\cfe < +1.7$ \citep[$\feh = -4.15$;][]{hansen2014}.

\begin{figure}[pt!]
\centering
\includegraphics[width=1.0\columnwidth]{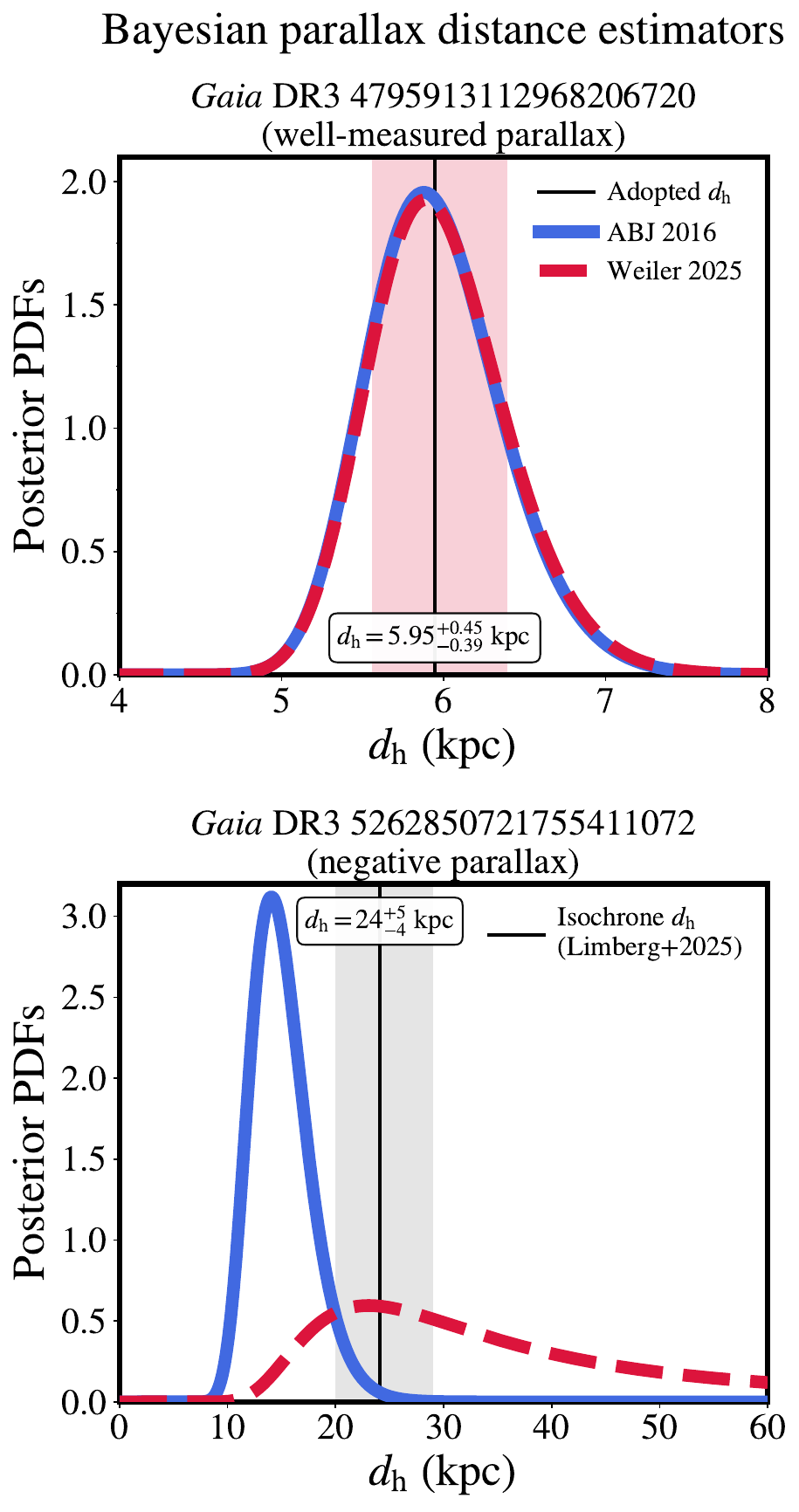}
\caption{Distance PDFs for both Bayesian estimators; exponentially decreasing space density prior \citep[blue;][]{abj2016} and \citet[][red dashed lines]{weiler2025plx}. Top panel: \gdr (this work). Our adopted $d_{\rm h} = 5.95^{+0.45}_{-0.39}\,{\rm kpc}$ value is marked with a black line and the red shaded area. Bottom panel: UMP star GDR3\_526285 with negative parallax \citep[][]{limberg2025ump_gaiaxp, ji2026natas}. The black line and gray shaded region depict the isochrone-based heliocentric distance for GDR3\_526285 by \citet[][]{limberg2025ump_gaiaxp}.
\label{fig:dist}}
\end{figure}

With both our UMP star compilation and the updated \texttt{JINAbase} collection at hand, we calculate corrections for the evolutionary carbon depletion in the RGB due to CN cycling \citep[][]{Stancliffe2009, placco2014Carbon}. To automate this task, we develop the \texttt{carbcor} Python package \citep{carbcor}\footnote{\url{https://github.com/guilimberg/carbcor}.}, which takes input \logg, [Fe/H], and measured [C/Fe] from a list of stars and fetches the corrections from the online tool provided by \citet[][]{placco2014Carbon}. Appendix \ref{carbcor} contains relevant information about installation and usage of \texttt{carbcor} for analyses of both individual stars and large samples.
    
\section{Methods} \label{sec:methods}

\subsection{Distance} \label{subsec:dist}

Due to the fact that \gdr has such a low metallicity, it is bluer and more luminous than its more metal-rich counterparts at the same evolutionary stage. \citet[][]{limberg2025ump_gaiaxp} demonstrated that, for this reason, widely-used photo-geometric methods in the literature underestimate the distance of another UMP star GDR3\_526285 by a factor of $>$30\% (e.g., \citealt{BailerJones2021gaiaEDR3} and \citealt{Anders2022starhorseEDR3}, the latter with the \texttt{StarHorse} code; \citealt{Queiroz2018}). Given the high quality of \gdr's astrometry in \textit{Gaia} DR3 ($\varpi_{\rm cor}/\sigma_\varpi > 14$, where $\sigma_\varpi$ is the parallax uncertainty after error inflation; see Table \ref{tabelao}), we derive our own purely-geometric distance for this UMP star.

We experiment with two different Bayesian estimators to convert \textit{Gaia} DR3 parallax measurements to heliocentric distances ($d_{\rm h}$) for \gdr. \citet[][]{abj2016} implement an exponentially decreasing space density prior \citep[][]{bailerjones2015edsd}, which has been extensively employed for \textit{Gaia} data \citep[e.g.,][]{bailerjones2018gaiaDR2, BailerJones2021gaiaEDR3}. On the other hand, \citet[][]{weiler2025plx} simply enforces a prior probability of zero for negative parallax values and a uniform prior otherwise. Since \gdr's parallax is remarkably well-measured, we expect these approaches to converge to the same $d_{\rm h}$. Hence, to establish which is really more appropriate for usage with UMP stars, we also compute posterior probability density functions (PDFs) with both methods for GDR3\_526285, which has a negative parallax in the \textit{Gaia} DR3 catalog. 

The results for the above-described exercise are summarized in Figure \ref{fig:dist}. Indeed, for \gdr, both Bayesian distance estimators result in essentially identical $d_{\rm h}$ estimates of $\sim$6\,kpc (top panel in Figure \ref{fig:dist}), as expected given the well-resolved parallax measurement. However, in the case of GDR3\_526285, the two methods provide drastically different PDFs. For the \citet[][]{abj2016} approach, we found $d_{\rm h} = 15.0 \pm 2.7\,{\rm kpc}$; this determination is underestimated by a factor of $\sim$35\% in comparison to the isochrone-based distance of $\sim$24$\pm$5\,kpc for the same star \citep[][]{limberg2025ump_gaiaxp}. On the other hand, the \citet[][]{weiler2025plx}\footnote{\url{https://github.com/fjaellet/weiler2025}.} PDF peaks at $d_{\rm h} \sim 23\,{\rm kpc}$, which is consistent with the \citet[][]{limberg2025ump_gaiaxp} value. We do note the extended tail toward larger distances in the \citet[][]{weiler2025plx} PDF, which translates to a median of $d_{\rm h} = 35.7^{+49.1}_{-14.2}\,{\rm kpc}$, where lower/upper bound uncertainties are the 16$^{\rm th}$/84$^{\rm th}$ percentiles.

The takeaway from our exploration with Bayesian parallax distances is that the \citet[][]{weiler2025plx} approach is more appropriate when parallax is negative and/or uncertainties are large. For negative parallax measurements, the \citet[][]{abj2016} method produces a PDF that substantially underestimates the `true' (isochrone) distance. Perhaps even more concerning, the extracted $d_{\rm h}$ estimate from this PDF has unrealistically small uncertainties, 
making it completely incompatible with the true value. On the contrary, the \citet[][]{weiler2025plx} estimator, despite not being as precise, does not provide a misleadingly well-resolved PDF and even peaks at the `correct' $d_{\rm h}$. Although our investigation is certainly not exhaustive, we expect it to serve as a starting point for future analyses of UMP stars with parallax-based distances. With all that said, the final median distance for \gdr derived from the \citet[][]{weiler2025plx} PDF is $d_{\rm h} = 5.95^{+0.45}_{-0.39}\,{\rm kpc}$, where lower/upper bound uncertainties are 16$^{\rm th}$/84$^{\rm th}$ percentiles from the same distribution. Distances from both \citet[][]{BailerJones2021gaiaEDR3} and \citet[][\texttt{StarHorse}]{Anders2022starhorseEDR3} are also displayed in Table \ref{tabelao}.

\begin{figure*}[pt!]
\centering
\includegraphics[width=2.1\columnwidth]{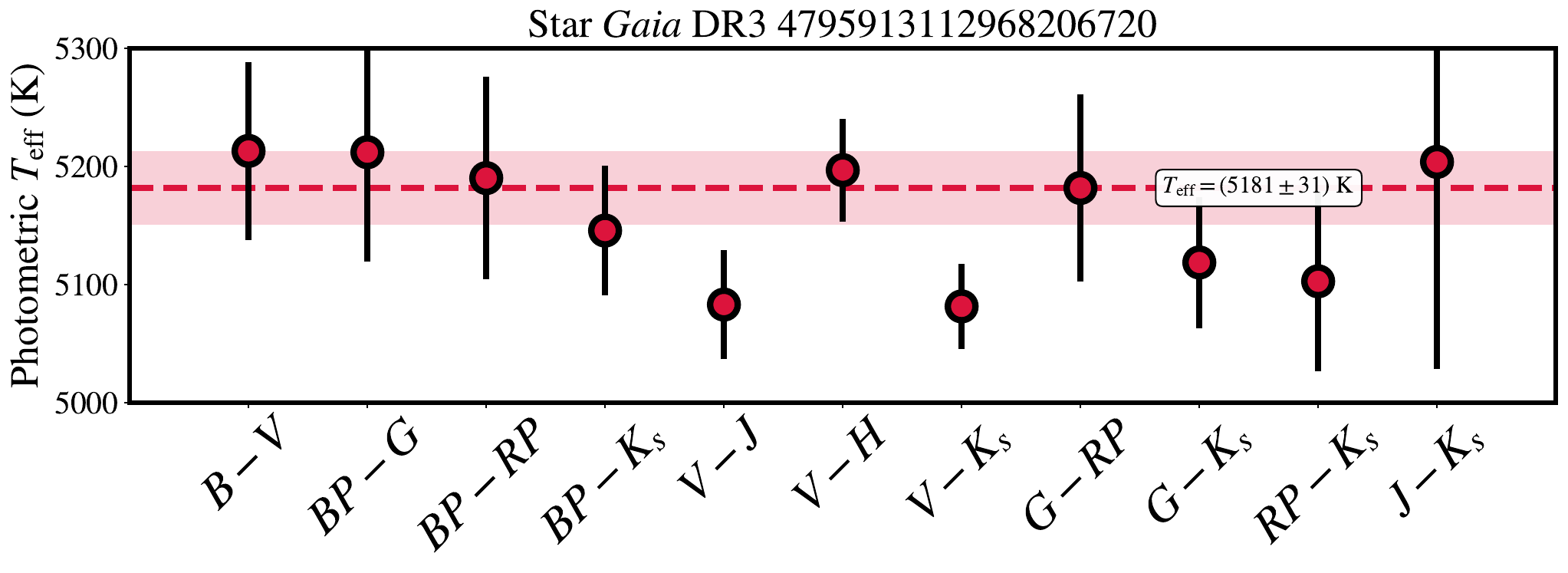}
\includegraphics[width=2.1\columnwidth]{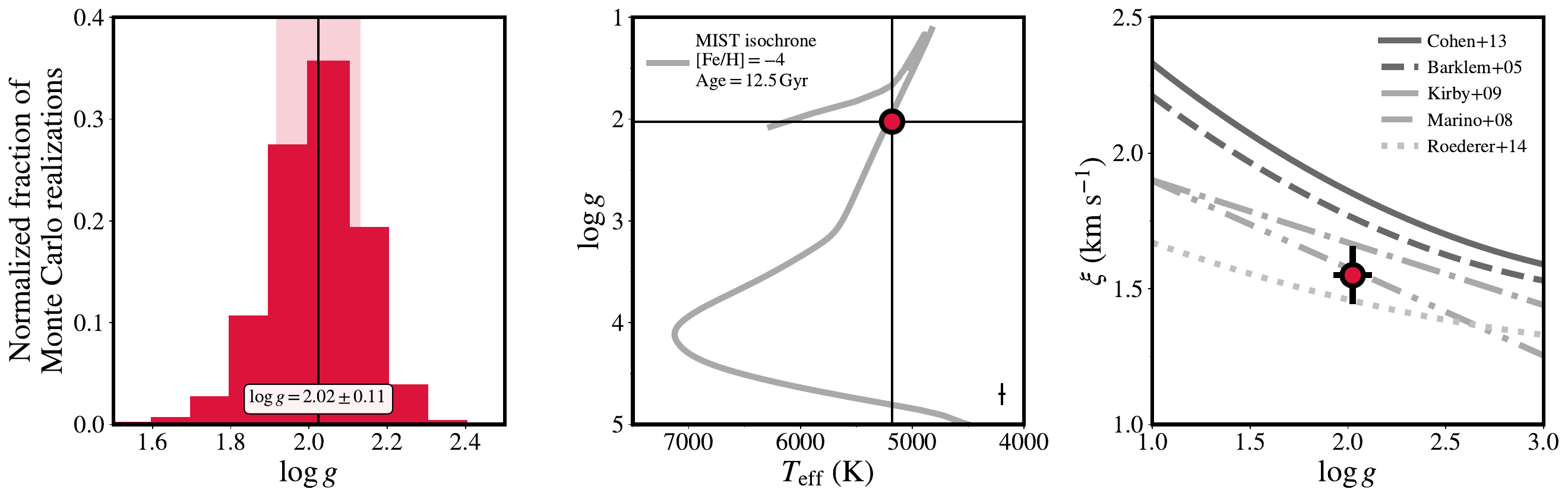}
\caption{Top panel: photometric \teff values for \gdr from different color--\teff--\feh calibrations based on \citet[][JKC$+$2MASS]{GHB2009colorTeffcalib} and \citet[][\textit{Gaia}$+$2MASS]{mucciarelli2021calib}. The various \teff estimates are organized from the bluest to the reddest bands in the first terms of the color combinations (red symbols with black edges and black error bars). From left to right: $B-V$ ($5212\pm 75$\,K), ${BP}-G$ ($5212\pm 93$\,K), ${BP}-{RP}$ ($5190\pm 86$\,K), ${BP}-K_s$ ($5146\pm 55$\,K), ${V}-{J}$ ($5084\pm 46$\,K), ${V}-{H}$ ($5198\pm 43$\,K), ${V}-{K_s}$ ($5081\pm 36$\,K), ${G}-{RP}$ ($5181\pm 79$\,K), ${G}-{K_s}$ ($5118\pm 56$\,K), ${RP}-{K_s}$ ($5102\pm 76$\,K), and ${J}-{K_s}$ ($5201\pm 175$\,K). The final adopted $\teff = 5181 \pm 31$\,K for \gdr is marked as the red dashed line and shaded region. Bottom left: $\logg$ distribution resulting from our Monte Carlo sampling method with \gdr's adopted $\logg = 2.02\pm0.11$ marked as the black line and shaded region. Bottom middle: \teff versus \logg for a solar-scaled MIST isochrone at $\feh = -4$ and ${\rm age} = 12.5\,{\rm Gyr}$ (gray line). Sizes of error bars for \gdr are shown in the bottom right corner if this panel. Bottom right: \logg versus \vturb. Different gray lines are fits to this relation, mostly by \citet[][]{AlexJi2023ret2}, for different data sets of metal-poor stars; \citet[][solid]{Cohen2013}, \citet[][dashed]{barklem2005}, \citet[][dash-dot]{Kirby2009sculptor}, \citet[][dash-double dot]{Marino2008m4}, and \citet[][dotted]{roederer2014}.
\label{stellarparams}}
\end{figure*}


\subsection{Stellar parameters} \label{params}

We combine multi-band photometry, our distance estimate, and MIKE/Magellan spectroscopy to obtain accurate and precise stellar parameters for \gdr. We apply RGB-specific color--\teff--\feh relations from both \citet[][]{GHB2009colorTeffcalib} and \citet[][]{mucciarelli2021calib} using JKC$+$2MASS and \textit{Gaia}$+$2MASS color combinations, respectively. We consider dereddened colors only, following Section \ref{photometry_astrometry}, and always assume $\feh = -4$, which is formally at the edge of validity of the fitting functions. We derive a total of 11 photometric \teff values as displayed in the top panel of Figure \ref{stellarparams}. The individual uncertainties are computed from the resulting Gaussian distributions out of $10^4$ Monte Carlo realizations of each \teff value. These statistical errors are added in quadrature with the reported systematic scatters from the original color--\teff--\feh calibrations. Our final adopted $\teff = 5181\pm31$\,K for \gdr was taken as the median of these various determinations, with the median absolute deviation as the uncertainty.

For \logg, we utilize the fundamental relation between this quantity, \teff, and luminosity:
\begin{linenomath}  
\begin{dmath} \label{eq4}
    \logg = \logg_\odot + \log{\left( \dfrac{M}{M_\odot} \right)} + 4 \log{\left(\dfrac{\teff}{T_{{\rm eff},\odot}}\right)} + 0.4(M_{\rm bol} - M_{\rm bol, \odot}),
\label{distlogg}
\end{dmath}
\end{linenomath}
where $\logg_\odot = 4.44$, $T_{{\rm eff},\odot} = 5772$\,K, and $M_{\rm bol, \odot} = 4.74$ are the solar effective temperature, logarithm of surface gravity, and bolometric magnitude values recommended by the International Astronomical Union 2015 resolution \citep[][]{Prsa2016_IAUsolarValues}. The assumed mass of \gdr is $M = 0.75\pm0.15\,\msun$, which is typical for old metal-poor RGB stars, and
\begin{equation} \label{eq5}
    M_{\rm bol} = {\rm BC}(m_\lambda) + m_{\lambda,0} - \mu
\end{equation}
is the bolometric magnitude of \gdr, where ${\rm BC}(m_\lambda)$ is the \citet[][]{CasagrandeVandenBerg2018gaia} bolometric correction in an $m_\lambda$ filter; we utilize \textit{Gaia} $G$ for $m_\lambda$. The distance modulus is defined as 
\begin{equation} \label{eq6}
    \mu = 5(\log{d_{\rm h}}-1),
\end{equation}
where the heliocentric distance $d_{\rm h}$, in units of parsec, is taken from our Bayesian parallax-based method in Section \ref{subsec:dist}.

We develop a full Monte Carlo approach to derive the final \logg for \gdr and propagate uncertainties. We implement an initial 9-realization loop starting at $\logg = 2.0$ and, for each iteration, re-sample all quantities in Equations \ref{eq4}, \ref{eq5}, and \ref{eq6} according to their observational errors. For ${\rm BC}(G)$, we fix $\feh = -4$ and vary \logg as it is updated within these initial loops. We then take the converged bolometric correction value and its dispersion as inputs for a final sampling with $10^6$ realizations. Performing the calculation in this way, instead of just re-sampling everything all at once, is necessary since the bolometric correction estimations are computationally expensive. The adopted $\logg = 2.02 \pm 0.11$ for \gdr is the mean and standard deviation of the resulting distribution (bottom left panel in Figure \ref{stellarparams}). We compare our stellar parameters with the expected \teff and \logg for a solar-scaled model isochrone ($\feh = -4$ and ${\rm age} = 12.5\,{\rm Gyr}$) from MESA\footnote{Modules for Experiments in Stellar Astrophysics \citep[][]{paxton2011_mesa}.} Isochrones and Stellar Tracks \citep[MIST;][]{MESA_0, MESA_1} and find an excellent agreement with an RGB solution (bottom middle panel in Figure \ref{stellarparams}). The expected mass from this MIST isochrone for \gdr's \teff is 0.81\,\msun, which would change the resulting \logg value from Equation \ref{eq4} by $+$0.03\,dex, hence well within our statistical uncertainty.


The iron-abundance metallicity for \gdr ($\feh=-4.06\pm0.17$) is determined spectroscopically from the equivalent width (EW) analysis of \ion{Fe}{1} absorption lines in the MIKE spectrum by fixing \teff and \logg. Our determination is 1\,$\sigma$ consistent with the $\feh = -4.09$ value reported by \citet[][]{yong2021_SMSSmetalpoor}. The EW values were obtained by fitting Gaussian profiles to the observed absorption lines using \texttt{smhr} \citep[][]{smhr}. 34 \ion{Fe}{1} and 2 \ion{Fe}{2} lines were measured. \texttt{smhr} determines \feh using the 2017 version of the \texttt{MOOG}\footnote{\href{https://github.com/alexji/moog17scat}{https://github.com/alexji/moog17scat}} radiative-transfer code \citep{Sneden1973moog,sobeck2011}, employing 1D plane-parallel model atmospheres with no overshooting \citep{Castelli2003atmospheres} and assuming local thermodynamic equilibrium (LTE).



\setlength{\tabcolsep}{0.4em}
\begin{deluxetable}{lrrrrrr}[!ht]
\tabletypesize{\footnotesize}
\tablewidth{0pc}
\tablecaption{LTE Abundances for Individual Species \label{table:abund}}
\tablehead{
\colhead{Ion}                         & 
\colhead{$\log\epsilon_{\odot}$\,(X)} & 
\colhead{$\log\epsilon$\,(X)}         & 
\colhead{$\mbox{[X/Fe]}$}             & 
\colhead{$\Delta$NLTE\xx}             & 
\colhead{$\sigma_{\rm tot}$\yy}       &
\colhead{$N$}                         }
\startdata
C           &    8.43 &    5.90 &    1.53 & \nodata & 0.19 &   2 \\ 
C\vv        &    8.43 &    5.91 &    1.54 & \nodata & 0.18 &   2 \\ 
N           &    7.83 &    6.41 &    2.64 & \nodata & 0.28 &   2 \\ 
\ion{O}{1}  &    8.69 &    7.25 &    2.62 & $-$0.06 & 0.18 &   3 \\ 
\ion{Na}{1} &    6.24 &    3.78 &    1.60 & $-$0.49 & 0.18 &   2 \\ 
\ion{Mg}{1} &    7.60 &    4.88 &    1.34 &    0.20 & 0.13 &   8 \\ 
\ion{Al}{1} &    6.45 &    2.21 & $-$0.18 &    0.80 & 0.18 &   2 \\ 
\ion{Si}{1} &    7.51 &    4.24 &    0.79 &    0.11 & 0.23 &   1 \\ 
\ion{Ca}{1} &    6.34 &    2.44 &    0.16 &    0.35 & 0.21 &   1 \\ 
\ion{Sc}{2} &    3.15 & $-$0.82 &    0.09 & \nodata & 0.10 &   4 \\ 
\ion{Ti}{2} &    4.95 &    1.19 &    0.30 &    0.10 & 0.13 &  10 \\ 
\ion{Cr}{1} &    5.64 &    1.17 & $-$0.41 &    0.91 & 0.10 &   2 \\ 
\ion{Mn}{1} &    5.43 &    0.58 & $-$0.79 &    1.00 & 0.18 &   2 \\ 
\ion{Fe}{1} &    7.50 &    3.44 &    0.00 &    0.41 & 0.17 &  34 \\ 
\ion{Fe}{2} &    7.50 &    3.42 & $-$0.03 &    0.02 & 0.08 &   2 \\ 
\ion{Co}{1} &    4.99 &    1.31 &    0.38 & \nodata & 0.11 &   2 \\ 
\ion{Ni}{1} &    6.22 &    2.70 &    0.54 & \nodata & 0.24 &   4 \\ 
\ion{Sr}{1} &    2.87 & $-$1.83 & $-$0.64 & \nodata & 0.15 &   2 \\ 
\ion{Ba}{1} &    2.18 & $-$2.42 & $-$0.54 & \nodata & 0.15 &   1 \\ 
\enddata
\tablenotetext{a}{Evolutionary corrections from \citet{placco2014Carbon}.}
\tablenotetext{b}{Assuming $\feh=-4.0$ for the \ion{Na}{1} correction calculation.}
\tablenotetext{c}{Calculated from the quadratic sum of individual error estimates. See Table~\ref{sys}.}
\tablerefs{
NLTE corrections -- 
\ion{Na}{1}: \citet{lind2011};
\ion{Mg}{1}: \citet{bergemann2015};
\ion{Al}{1}: \citet{nordlander2017b};
\ion{Si}{1}: \citet{bergemann2013};
\ion{Ca}{1}: \citet{mashonkina2007};
\ion{Ti}{2}: \citet{bergemann2011};
\ion{Cr}{1}: \citet{bergemann2010jinaB};
\ion{Mn}{1}: \citet{bergemann2019};
\ion{Fe}{1} and \ion{Fe}{2}: \citet{bergemann2012b}.
}
\end{deluxetable}


With our \teff, \logg, and \feh values fixed, we proceed to estimate the microturbulence velocity (\vturb) for \gdr. To do so, we enforce the balance between \ion{Fe}{1} abundances and their reduced equivalent widths
. The $\vturb = 1.55\pm 0.15\,\kmsec$ we obtain (bottom right panel in Figure \ref{stellarparams}) is well within the \logg--\vturb tracks from previous analyses of metal-poor stars \citep[][]{Marino2008m4, Kirby2009sculptor, roederer2014}. We note that, at the estimated \logg of \gdr, the fitted relations for \citet[][]{barklem2005} and \citet[][]{Cohen2013} data sets predict $\sim$0.20\,\kmsec higher \vturb, consistent with the 1.8\,\kmsec value from \citet{yong2021_SMSSmetalpoor}, which would reduce the final \feh by $\sim$0.05\,dex. All stellar atmospheric parameters are listed in Table \ref{tabelao}.

\begin{figure*}[pt!]
\centering
\includegraphics[width=2.1\columnwidth]{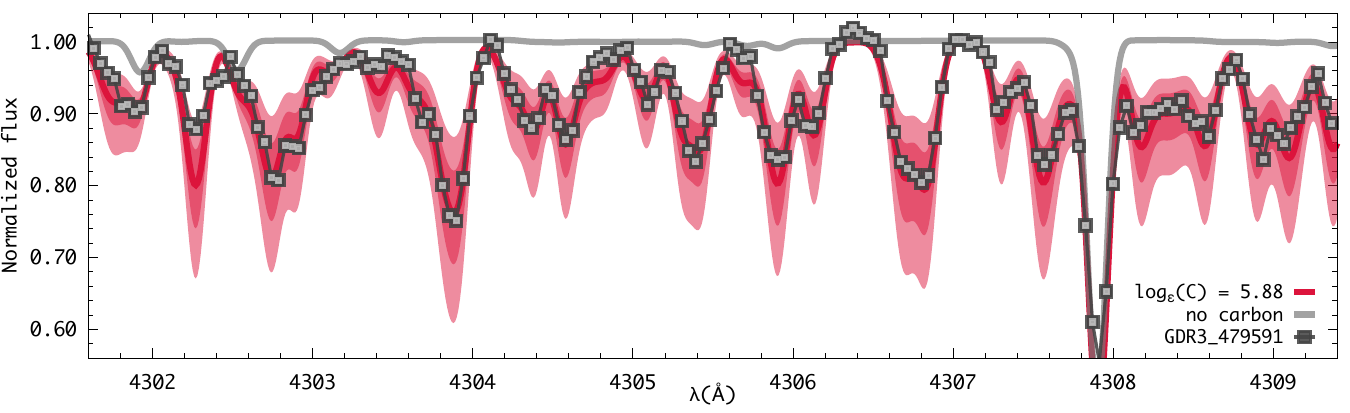}
\includegraphics[width=2.1\columnwidth]{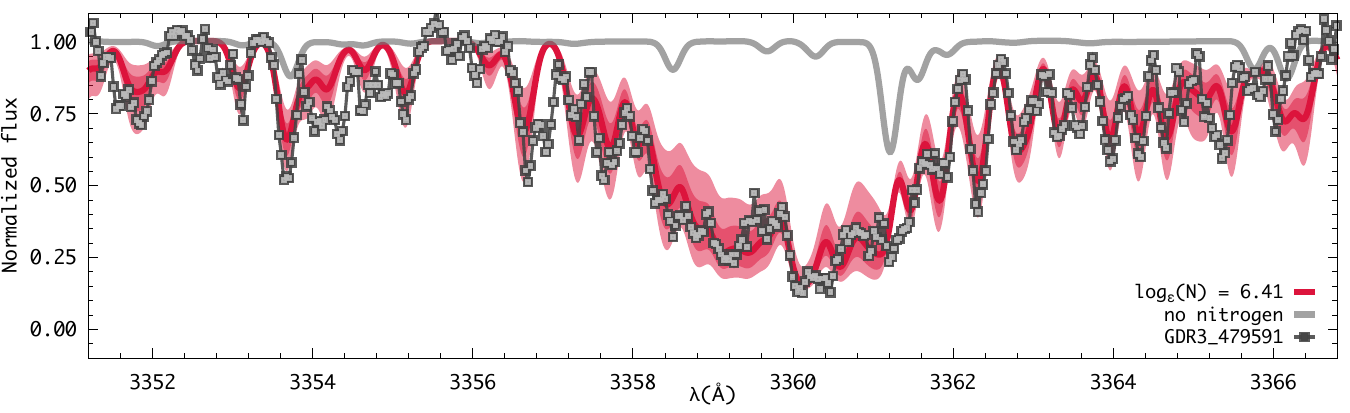}
\includegraphics[width=2.1\columnwidth]{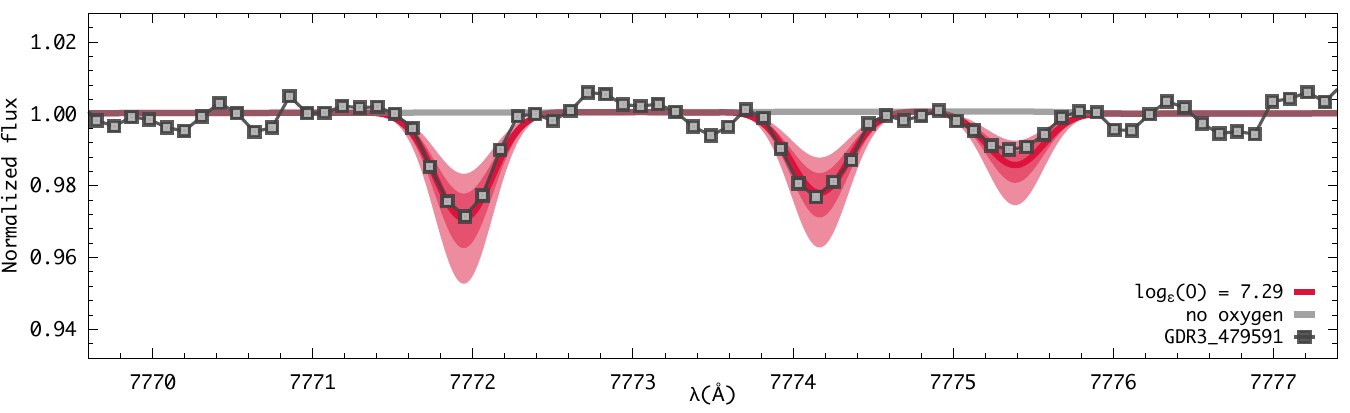}
\caption{Abundance determination via spectral synthesis for carbon (top), nitrogen (middle), and oxygen (bottom). The filled gray squares connected by the black line represent the MIKE spectrum, the crimson red line is the best fit, and the shaded regions represent $\pm0.15$/$\pm0.30$~dex (C and O) and $\pm0.20$/$\pm0.40$~dex (N) from the best-fit abundance. Also shown are synthetic spectra without carbon, nitrogen, and oxygen (gray lines).
\label{fig:spec}}
\end{figure*}

\subsection{Chemical abundances} \label{abunds}


Chemical abundances for 17 elements are determined for \gdr using EWs and spectral synthesis. The atomic and molecular line lists were generated with the \texttt{linemake} code\footnote{\href{https://github.com/vmplacco/linemake}{https://github.com/vmplacco/linemake/}} \citep{Placco2021linemake}. Logarithmic number abundances ($\log\epsilon$(X)) and abundance ratios (\xfe{X}) adopt the solar photospheric chemical composition 
from \citet{asplund2009}. Table~\ref{table:abund} lists the average abundance values and the number of lines measured ($N$) for each element. The full line-by-line abundances as well as atomic and molecular data are listed in Appendix \ref{appendix:atomic_data}. The $\sigma_{\rm tot}$ values represent the total uncertainty budget. For abundances determined from EWs, the uncertainty is represented by the standard deviation across different lines of the same element. For the spectral synthesis, uncertainties were estimated by minimizing the residuals between the MIKE data and a set of synthetic spectra. We provide complete systematic abundance errors in Appendix \ref{appendix:sys_uncs}.

The CH~G-band is synthesized to determine the carbon abundance. We find $\eps{C}=5.88$ in the 4300--4315\,\angs region and $\eps{C}=5.92$ in the 4320--4326\,\angs region, with an average abundance of $\eps{C}=5.90$. A fixed $^{12}{\rm C}/^{13}{\rm C}=4$ is assumed since the carbon features most sensitive to isotopic ratio changes at 4217--4220\,\angs are too weak. This carbon-abundance value is consistent with the \citet{yong2021_SMSSmetalpoor} determination of $\eps{C} = 5.86$ for \gdr. The top panel of Figure~\ref{fig:spec} shows the spectral synthesis of a portion of the CH molecular feature. The filled gray squares connected by the black line represent the MIKE spectrum, the crimson red line is the best fit, and the shaded regions represent $\pm0.15$\,dex and $\pm0.30$\,dex from the best-fit abundance. The gray line shows a synthetic spectrum without carbon. 

The nitrogen abundance is determined from two molecular features, NH at 3360\,\angs and CN at 3883\,\angs (using a fixed $\eps{C}=5.90$), both with a best-fit abundance of $\eps{N}=6.41$. The spectral synthesis for the NH feature is shown in the middle panel of Figure~\ref{fig:spec}, where the shaded areas represent $\pm0.20$ and $\pm0.40$~dex from the best-fit value. For oxygen, the three atomic \ion{O}{1} features at 7771\,\angs ($\eps{O}=7.29$), 7774\,\angs ($\eps{O}=7.29$), and 7775\,\angs ($\eps{O}=7.17$) are synthesized. The best-fit abundance is shown in the bottom panel of Figure~\ref{fig:spec}, where the shaded areas represent $\pm0.15$ and $\pm0.30$~dex. While our nitrogen abundance for \gdr is consistent with the value reported by \citet[][$\eps{N} = 6.35$]{yong2021_SMSSmetalpoor}, ours is the first oxygen detection in this UMP star.

\begin{figure*}[pt!]
\centering
\includegraphics[width=2.1\columnwidth]{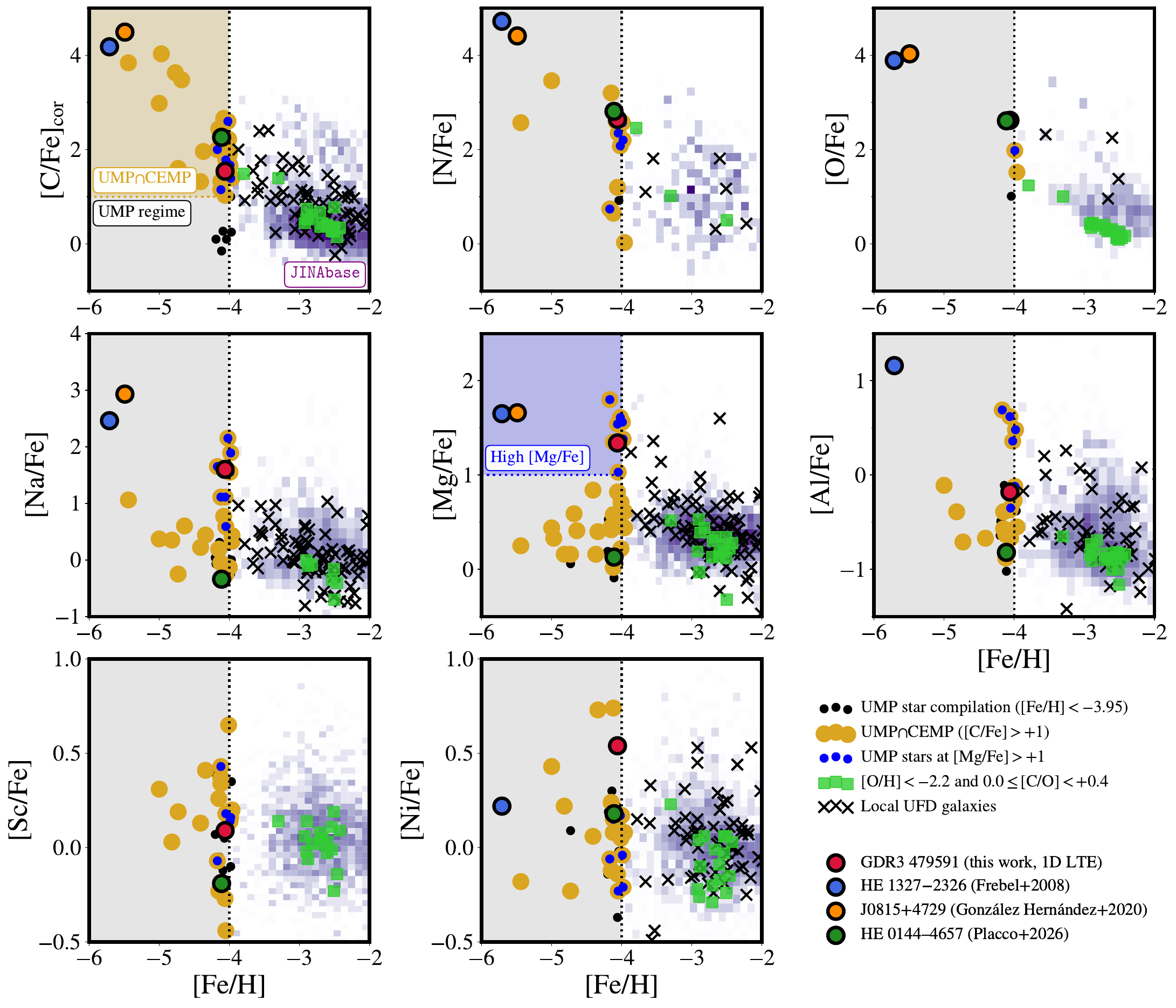}
\caption{Various chemical abundance spaces; [Fe/H] versus [X/Fe], where `X' represents relevant elements for parameter recovery in Population~III supernova nucleosynthesis yield fitting. Top row: C, N, and O (from left to right column). All carbon abundances are corrected (`cor' subscript) for evolutionary depletion in the RGB following \citet[][]{placco2014Carbon} using the \texttt{carbcor} package (see text). We highlight the CEMP region ($\cfe \geq +1$) with a yellow background. Middle row: Na, Mg, and Al. The \textit{locus} occupied by Mg-rich stars ($\rm[Mg/Fe] \geq +1$) is colored blue. Bottom row: Sc and Ni. In all panels, the CNO-enhanced UMP stars (1D~LTE only) are displayed with larger symbols with black edges; \gdr (this work, red circles), HE~1327$-$2326 \citep[][blue]{frebel2008}, J0815$+$4729 \citep[][orange]{GonzalezHernandez2020}, and HE~0144$-$4657 \citep[][green]{placco2026umpHelmiStream}. The UMP stars in our compilation (Section \ref{subsec:lit_ump}/Appendix \ref{appendix:ump_stars}/Table \ref{tab:ump_comp}) are shown as the black dots. Those UMP stars that are also CEMP are plotted as the larger yellow dots and the high-Mg ones exhibit a smaller blue dot on top of them. Stars in \texttt{JINAbase} \citep[][2021 update]{jina} with abundances similar to the LAP1-B galaxy \citep[redshift $z\sim6.6$;][]{nakajima2025pop3Gal} are marked as the green squares. Stars from Milky Way UFD satellites are the black crosses (all references in Appendix \ref{appendix:ufd_stars}). The complete \texttt{JINAbase} at $-4 < \feh \leq -2$ is presented as the purple 2D histogram (Appendix \ref{appendix:refs_jinabase}). 
\label{fig:all_abunds}}
\end{figure*}

Abundances for \ion{Al}{1} (3944\,\angs and 3961\,\AA), \ion{Si}{1} (3905\,\AA), \ion{Ca}{1} (4226\,\AA), \ion{Mn}{1} (4030\,\angs and 4033\,\AA), \ion{Sr}{2} (4077\,\angs and 4215\,\AA), and \ion{Ba}{2} (4554\,\AA) are determined through spectral synthesis, and, for the remaining elements listed in Table~\ref{table:abund}, we use EW curve-of-growth analysis. In particular, we find consistent measurements for both \ion{Na}{1}~D components and also for eight \ion{Mg}{1} lines. These are the first detections of both Si and Mn in \gdr. For other elements, our values are 1\,$\sigma$ with \citet{yong2021_SMSSmetalpoor}. The final average LTE abundances in \gdr for selected elements relevant for the recovery of Population~III progenitor parameters \citep[e.g.,][]{jarvis2026_SNyieldFitting} are displayed in Figure \ref{fig:all_abunds}.

Corrections for NLTE effects are obtained for nine elements using the INSPECT\footnote{\href{http://www.inspect-stars.com/}{http://www.inspect-stars.com/}} database (\ion{Na}{1}), \citet{nordlander2017b} (\ion{Al}{1}), and the MPIA NLTE\footnote{\href{https://nlte.mpia.de/}{https://nlte.mpia.de/}} \citep{mpia} database (\ion{Mg}{1}, \ion{Si}{1}, \ion{Ca}{1}, \ion{Ti}{2}, \ion{Cr}{1}, \ion{Mn}{1}, \ion{Fe}{1}, and \ion{Fe}{2}). For \ion{O}{1}, we employ the \texttt{TSFitPy}\footnote{\url{https://github.com/TSFitPy-developers/TSFitPy}.} NLTE Python wrapper \citep[][]{gerber2023_tsfitpy, storm2023_yttrium_tsfitpy} of \texttt{Turbospectrum} spectrum synthesis code \citep[][]{AlvarezPlez1998_Turbospectrum, Plez2012_Turbospectrum}, with model atom from \citet[][]{bergemann2021oxygenNLTEmodelAtom}. The average NLTE corrections are listed in the $\Delta{\rm NLTE}$ column of Table~\ref{table:abund}. Some species are heavily affected by NLTE effects, with the largest corrections for \ion{Mn}{1} ($+1.00$\,dex), \ion{Cr}{1} ($+0.91$\,dex), and \ion{Al}{1} ($+0.80$\,dex).

\subsection{Kinematics and dynamics} \label{subsec:kindyn}

We combine \textit{Gaia} DR3 astrometric observables, i.e., on-sky coordinates and proper motions, the geometric $d_{\rm h}$ from the \citet[][]{weiler2025plx} estimator, and MIKE \vlos to calculate kinematic parameters for \gdr. We convert these phase-space data to Galactocentric Cartesian positions and velocities using the \texttt{PyGaia}\footnote{\url{https://pypi.org/project/pygaia/}.} and \texttt{Astropy} toolkit  \citep[][]{astropy, astropy2018}. The adopted location of the Sun is $(X,Y,Z)_\odot = (-8.2, 0.0, 0.0)$, as recommended by \citet[][]{BlandHawthorn2016}, with a velocity vector $(V_x,V_y,V_z)_\odot = (11.10, 245.04, 7.25)\,\kmsec$, which includes the solar peculiar motion \citep[][]{schon2010} and the velocity of the local standard of rest \citep[$V_{\rm LSR} = 232.80\,\kmsec$;][]{mcmillan2017}. With this information at hand, we compute specific angular-momentum components $\mathbf{L} = (L_x, L_y, L_z)$ for \gdr, where $L_x = YV_z - ZV_y$, $L_y = ZV_x - XV_z$, and $L_z = XV_y - YV_x$. Note that, within this reference frame, $L_z < 0$ signifies prograde motion. For example, $L_{z,\odot} = -2009\,{\rm kpc}\,\kmsec$ for the Sun.

We propagate observational Gaussian uncertainties by re-sampling the Cartesian phase-space information of \gdr with $10^4$ realizations in a Monte Carlo scheme. For each of these, we compute the specific total orbital energy $E = \frac{1}{2}(V_x^2 + V_y^2 + V_z^2) + \Phi(\mathbf{r})$, where $\Phi$ represents the \citet[][]{mcmillan2017} axisymmetric Milky Way model potential (total mass $M_{\rm MW} = 1.3 \times 10^{12}\,\msun$) and $\mathbf{r} = (X,Y,Z)$ is the position vector of \gdr. We simultaneously integrate an orbit within the same static potential $\Phi$ for each realization by calculating \gdr's trajectory for 20\,Gyr forward using the \texttt{AGAMA} package \citep[][]{agama}. We take the medians of the resulting distributions for all kinematic and orbital parameters as our nominal values with 16$^{\rm th}$/84$^{\rm th}$ percentiles as lower/upper bound uncertainties. Table \ref{tabelao} lists derived values for several dynamical quantities of interest, including the aforementioned angular-momentum components ($L_x$, $L_y$, and $L_z$) and total orbital energy ($E$), but also orbital apocenter ($r_{\rm apo}$) and pericenter ($r_{\rm peri}$), orbital eccentricity $e = (r_{\rm apo} - r_{\rm peri})/(r_{\rm apo} + r_{\rm peri})$, and orbital inclination $\theta = \arccos{\left(L_z/L\right)}$, where $L = \sqrt{L_x^2 + L_y^2 + L_z^2}$)
.

\section{Results} \label{results}

\subsection{\gdr is a UMP star enhanced in several light elements} \label{lightelements}

From our 1D LTE abundance analysis, we observe that \gdr is a CEMP star at $\cfe = +1.54$ (Table \ref{table:abund}), after applying evolutionary correction for carbon depletion in the RGB \citep{placco2014Carbon}, which is typical of UMP stars (Section \ref{subsec:lit_ump}). Also, \gdr does not display over-enrichment in neutron-capture 
elements ($\rm[Sr/Fe] = -0.64$ and $\rm[Ba/Fe] = -0.54$), placing it in the category of so-called `CEMP-no' stars \citep[][]{beers2005}, i.e., its carbon enhancement is unrelated to binary mass transfer, but really depicts the abundance of the interstellar medium at the time and location of its birth \citep[e.g.,][]{Lucatello2005, Hansen2016cempNO}. Indeed, there is no evidence for \vlos variation in \gdr given our MIKE/Magellan, \citet{yong2021_SMSSmetalpoor}, and \textit{Gaia}~DR3's measurements (Table \ref{tabelao}). The most striking observations, however, are our detections of both nitrogen and oxygen in this star's atmosphere. We find that \gdr also exhibits excesses in both of these elements, namely $\rm[N/Fe] = +2.64$ and $\rm[O/Fe] = +2.62$. The resulting $\rm[C/N]=-1.10$ might indicate some degree of internal mixing \citep[e.g.,][]{Spite2005mix}.

From our curated list, we verify that only a few UMP stars have oxygen abundance measurements. The most extreme examples in Table \ref{tab:ump_comp} are HE~1327$-$2326 \citep[$\feh = -5.71$;][]{frebel2008} and J0815$+$4729 \citep[$\feh = -5.49$;][]{GonzalezHernandez2020}, both of which are CEMP UMP stars displaying $\rm[C{,}N{,}O/Fe] \gtrsim +4$. A more recent discovery is CEMP UMP star HE~0144$-$4657 ($\feh = -4.11$ and $\cfe = +2.26$, after evolutionary correction), which also exhibits both $\rm[N/Fe] = +2.81$ and $\rm[O/Fe] = +2.61$ enhancements \citep[][]{placco2026umpHelmiStream}, making it quite similar to \gdr. We refer to the collective formed by HE~1327$-$2326, J0815$+$4729, HE~0144$-$4657, and \gdr as `CNO-enhanced UMP stars' throughout this work, with the first two being the most notable objects within this class (lower \feh and higher $\rm[C{,}N{,}O/Fe]$). Other UMP stars with oxygen detections in our compilation include BPS~CS~22949$-$0037 ($\feh = -3.99$), BPS~CS~30336$-$0049 ($\feh = -4.04$), and BD$+$44~493 ($\feh = -3.96$) \citep[][respectively]{Cayrel2004, Lai2008_sn_yield_fits, placco2024}, but all at $\rm[O/Fe] < +2$.


The CNO enhancement of \gdr is accompanied by high element-to-iron ratios in other light species, namely $\rm[Na/Fe] = +1.11$, $\rm[Mg/Fe] = +1.54$, $\rm[Al/Fe] = +0.62$, and $\rm[Si/Fe] = +0.90$ (after NLTE corrections are accounted for; Table \ref{table:abund}). The over-abundance in these light elements is particularly noteworthy when compared to other UMP stars as well as the general low-metallicity \mw population from the updated \texttt{JINAbase} in the second row of Figure \ref{fig:all_abunds}. \citet[][]{aoki2018} had previously reported that $\sim$10\% of stars within $-4.5 < \feh < -3.5$ show $\rm[Mg/Fe] > +1$ \citep[see also][]{yoon2016}. Indeed, a non-negligible amount of UMP stars in our compilation, as well as some higher-\feh ones in \texttt{JINAbase}, do exhibit such high $\rm[Mg/Fe]$, including HE~1327$-$2326 \citep[][]{frebel2008} and J0815$+$4729 \citep[][]{GonzalezHernandez2020}. All of these high-$\rm[Mg/Fe]$ UMP stars are CEMP, and most of them, indeed, also display comparatively high $\rm[Na/Fe]$ and $\rm[Al/Fe]$. One exception is actually the \citet[][]{placco2026umpHelmiStream} HE~0144$-$4657 star, which is CNO-enhanced, but with $\sim$solar-level $\rm[Na{,}Mg{,}Al{,}Si/Fe]$ (even after $+$0.8\,dex NLTE correction for Al).

\begin{figure*}[pt!]
\centering
\includegraphics[width=1.36\columnwidth]{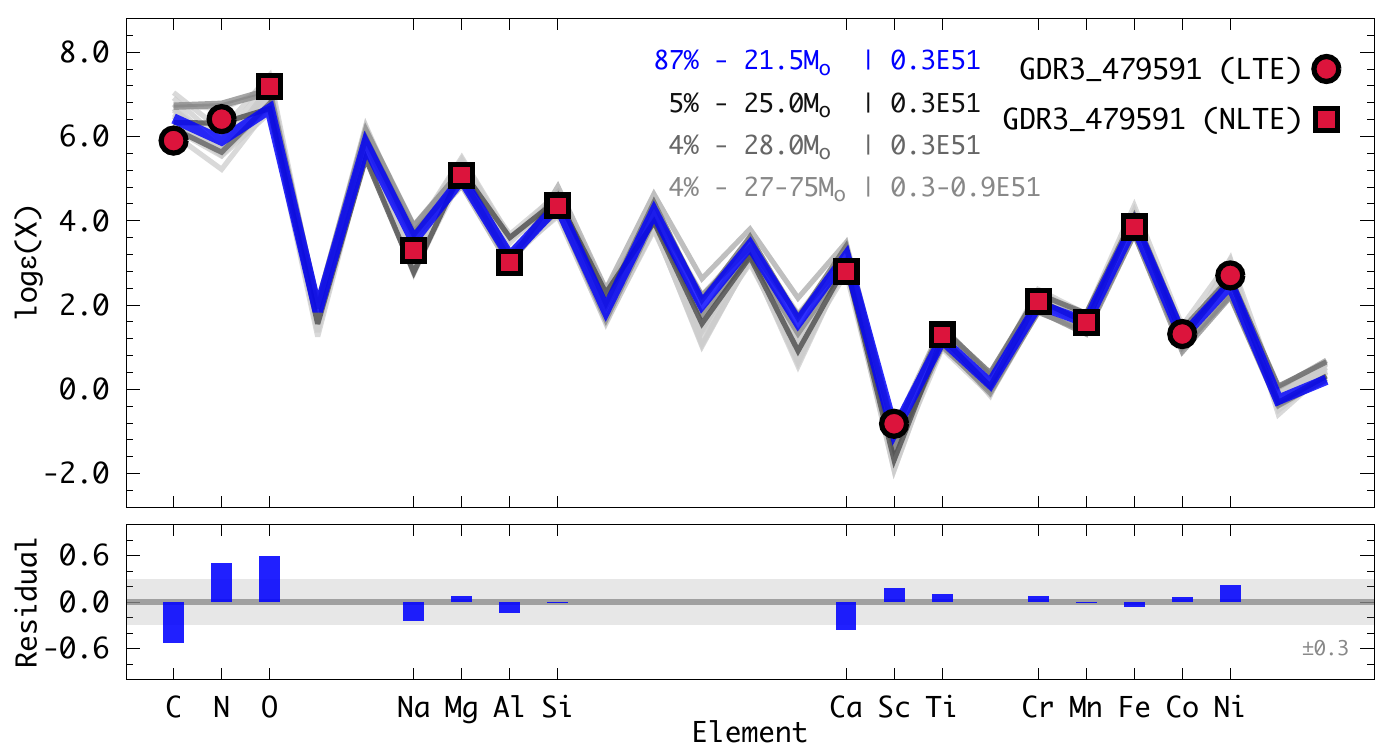}
\includegraphics[width=0.74\columnwidth]{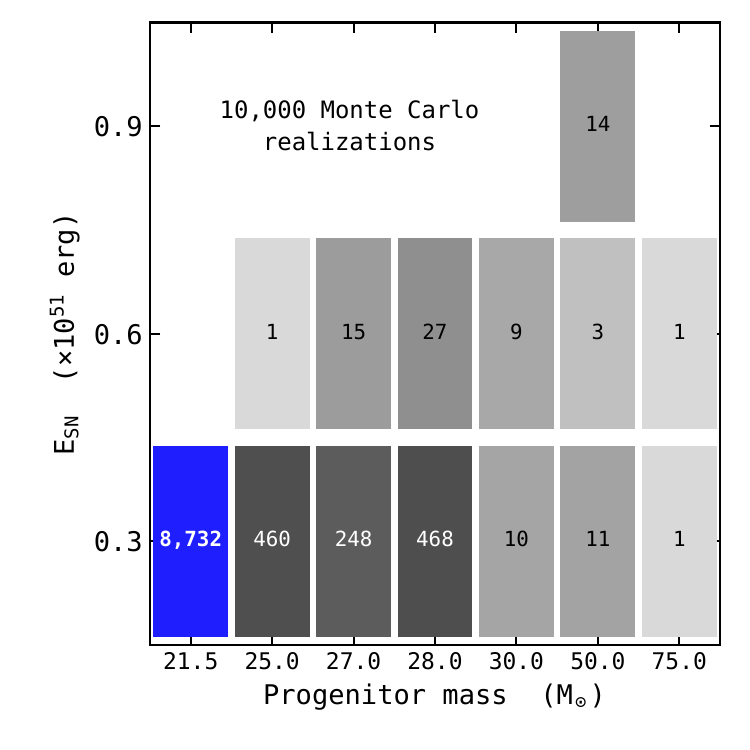}
\caption{Results for Population~III supernova yield fitting using \texttt{starfit} \citep[][]{HegerWoosley2002, Heger2010}. The left panel shows absolute abundances as a function of atomic number for the elements detected in \gdr, with residuals in the bottom, and the right panel shows a 2D histogram of Population~III progenitor mass versus explosion energy ($E_{\rm SN}$). The best-fit (87\% of realizations) abundance pattern is shown as the blue line in the left panel and the blue cell in the right panel (${\rm progenitor \ mass} = 21.5\,\msun$ and explosion energy $E_{\rm SN} = 0.3 \times 10^{51}\,{\rm erg}$) with other possible solutions in gray. The derived absolute elemental abundances $\log\epsilon$\,(X) for elements `X' in \gdr are the red symbols with black edges in the left panel, in 1D~LTE (circles) and NLTE (squares), when available.
\label{fig:starfit}}
\end{figure*}

The persistent over-enrichment across several light elements in \gdr is the telltale sign of the `mixing-and-fallback' process for supernovae explosions \citep{Umeda2003, ishigaki2014faintSN}. In this phenomenon, heavier elements that form within the inner regions of stellar interiors, such as Fe and other iron-peak elements, fall back into the remnant compact object after the supernova event. On the other hand, lighter elements (e.g., C, N, and O) located in the outer layers are released into the host galaxy's interstellar medium, even in the case of a low-energy explosion. In fact, UMP star HE~1327$-$2326 itself is one of the prototypes for this `faint' supernovae mechanism together with HE~0107$-$5240 \citep[$\feh = -5.44$ in Table \ref{tab:ump_comp};][]{iwamoto2005faintSN}, although the latter had no oxygen detection \citep[][]{Christlieb2004he0107-5240}. With \gdr, we have the opportunity to perform supernovae-yield fitting with a wide range of elements relevant for accurate parameter recovery \citep[e.g.,][]{jarvis2026_SNyieldFitting} and test the mixing-and-fallback predictions.


\subsection{The origin of light-element enhancement in \gdr in a low-energy Population~III supernovae progenitor} \label{sec:low-e_sn}

Being classified as a CEMP UMP object, \gdr is a primary candidate for being a genuine second-generation star, i.e., a direct descendant of a Population~III progenitor \citep[e.g.,][]{hartwig2018}. On top of that, oxygen is one of the most informative elements for distinguishing between mono- and multiply-enriched UMP stars \citep[][]{Hartwig2023}. Indeed, the high $\rm [O/Mg] = +1.28$ strongly indicates that \gdr was enriched by a single Population~III supernova. To test this hypothesis, we utilize the method described in \citet[][]{Hartwig2023}\footnote{\url{https://gitlab.com/thartwig/emu-c}.} to quantify the probability of \gdr being mono-enriched. We input abundances from Table \ref{table:abund}, NLTE-corrected when available and LTE otherwise, to find a probability of $91\pm9\%$ that \gdr is mono-enriched. 

With the above-described result, we proceed to fit the complete abundance pattern of \gdr with yields from Population~III supernova. For this exercise, we employ the \texttt{starfit}\footnote{\url{https://starfit.org/}.} tool, which includes Population~III yield models that account for the mixing-and-fallback mechanism. The range of values covered by these models is 9.6\,\msun to 100\,\msun for the Population~III progenitor mass, $(0.3 \ {\rm to} \ 10) \times 10^{51}\,{\rm erg}$ for the supernova explosion energy ($E_{\rm SN}$), and 0.0 to 2.5 for the mixing parameter \citep[][]{HegerWoosley2002, Heger2010}. We compute $10^4$ samples of \gdr's light-element abundances (${\rm atomic \ number} \leq 28$, i.e., excluding strontium and barium from Table \ref{table:abund}) assuming Gaussian uncertainties and NLTE corrections, when available. Figure \ref{fig:starfit} shows that the preferred progenitor (87\% of realizations) is a 21.5\,\msun Population~III star with $E_{\rm SN} = 0.3 \times 10^{51}\,{\rm erg}$ (blue line/cell in the left/right panel of Figure \ref{fig:starfit}), indicating a faint supernova origin for \gdr. Other possible solutions, with 25-to-28\,\msun progenitors, require explosion energies of $0.3\leq E_{\rm SN}/(10^{51}\,{\rm erg}) \leq 0.9$ (gray lines/cells in the left/right panel of Figure \ref{fig:starfit}).

Out of the collective of CNO-enhanced UMP stars, as defined in Section \ref{lightelements} above, their abundance patterns might also entail faint supernova progenitors. \citet[][]{GonzalezHernandez2020} report $0.3 \leq E_{\rm SN}/(10^{51}\,{\rm erg}) \leq 0.6$ for J0815$+$4729. On the other hand, the preferred solution ($\sim$80\% of realizations) for HE~0144$-$4657 is consistent with a more energetic core-collapse supernovae at $1.2 \leq E_{\rm SN}/(10^{51}\,{\rm erg}) \leq 1.5$ and only $<$20\% of realizations indicating a low-energy $E_{\rm SN} \leq 0.9 \times 10^{51}\,{\rm erg}$ explosion \citep[][]{placco2026umpHelmiStream}. In the case of HE~1327$-$2326, we use \texttt{starfit} and abundances from \citet{frebel2008} to estimate $E_{\rm SN} = 0.3 \times 10^{51}\,{\rm erg}$ for it. However, \citet[][]{ezzeddine2019he1327-2326} argued that only a `hypernova', with best-fit $E_{\rm SN} = 5 \times 10^{51}\,{\rm erg}$, can reproduce the measured $\rm[Zn/Fe]=+0.8$ in this star, reinforcing the importance of additional elements \citep[][]{jarvis2026_SNyieldFitting}.

\begin{figure}[pt!]
\centering
\includegraphics[width=1.0\columnwidth]{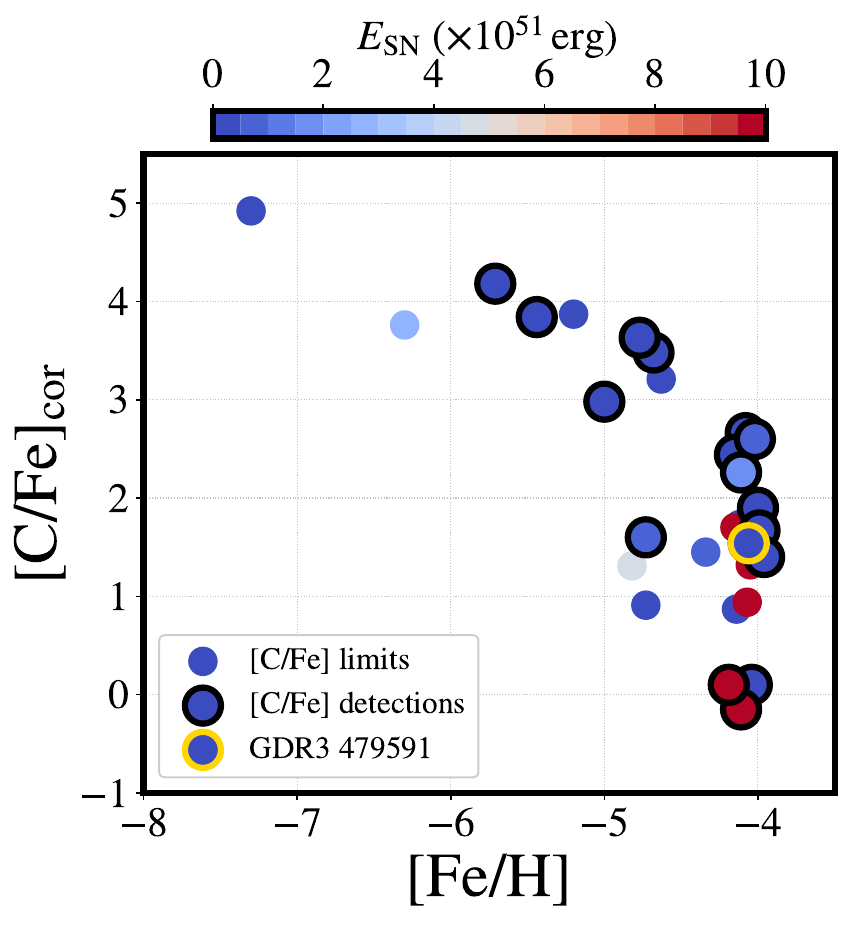}
\caption{\feh versus $\cfe_{\rm cor}$ for the UMP stars in our compilation (Section \ref{subsec:lit_ump}/Appendix \ref{appendix:ump_stars}/Table \ref{tab:ump_comp}), where the carbon-to-iron ratios are corrected (`cor' subscript) for evolutionary depletion in the RGB following \citet[][]{placco2014Carbon} using the \texttt{carbcor} package (see text). Different UMP stars are colored according to their associated best-fit Population~III progenitor explosion energies, from lower $E_{\rm SN}$ (blueish) to higher $E_{\rm SN}$ (reddish). \gdr is highlighted with a yellow edge.
\label{fig:cfe_pop3e}}
\end{figure}

We search the literature for Population~III supernova yield-fitting results for stars in our UMP compilation. All $E_{\rm SN}$ values are either from \citet[][]{placco2015seguefollowup} or from the references listed for each UMP star in Table \ref{tab:ump_comp}. Figure \ref{fig:cfe_pop3e} illustrates the $E_{\rm SN}$ dependence on carbon-to-iron ratio, where the majority of UMP stars in the CEMP regime at $\cfe > +1$ require low-energy supernova progenitors with $E_{\rm SN} \leq 0.9 \times 10^{51}\,{\rm erg}$. On the contrary, the hypernova case is usually favored for carbon-poor UMP stars such as those discovered in the Sculptor dwarf spheroidal galaxy \citep[][]{Skuladottir2021UMPsculptor} and the Large Magellanic Cloud \citep[][]{chiti2024lmc}, as well as the aforementioned GDR3\_526285 \citep[][and \citealt{limberg2025ump_gaiaxp}]{ji2026natas}. The discovery of \gdr and its enhancement in several light elements thus provide evidence for the faint supernova with mixing-and-fallback mechanism \citep[][]{Umeda2003, iwamoto2005faintSN} as the culprit for the CNO excess in the UMP regime, including the CEMP phenomenon. Nevertheless, to robustly assess its contribution to the chemical enrichment of ancient second-generation stars, abundance measurements of additional elements, or at least stringent upper limits, will certainly be required.


\subsection{The connection to early metal-poor galaxies: high~[C/O] does not translate to high [C/Fe]}

\begin{figure*}[pt!]
\centering
\includegraphics[width=2.1\columnwidth]{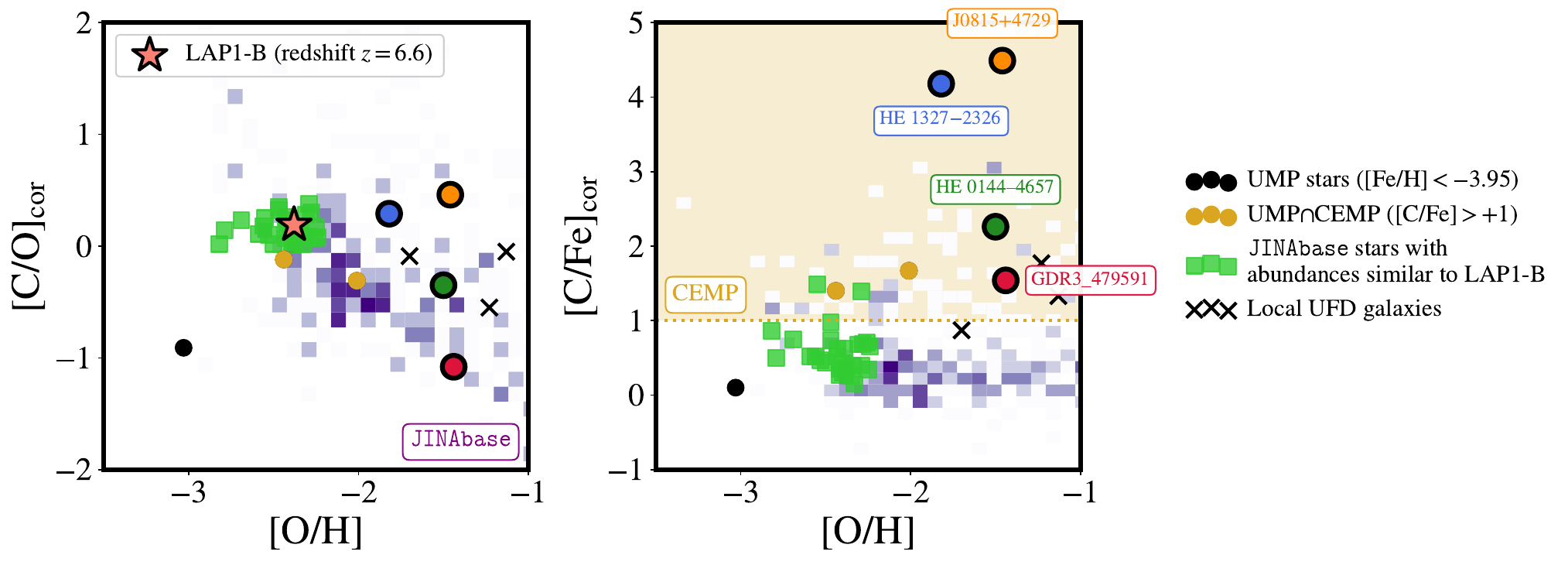}
\caption{Elemental abundance spaces relevant for direct comparisons with extragalactic sources, i.e., including oxygen detections. Left: $\rm[O/H]$ versus $\rm[C/O]_{\rm cor}$, where the stellar carbon abundances are corrected (`cor' subscript) for evolutionary depletion in the RGB following \citet[][]{placco2014Carbon} using the \texttt{carbcor} package (see text). The LAP1-B galaxy (redshift $z\sim6.6$) gas-phase abundances is marked as the star symbol \citep[][]{nakajima2025pop3Gal}. Right: $\rm[O/H]$ versus $\rm[C/Fe]_{\rm cor}$. The Stellar Archaeology-defined CEMP region ($\cfe \geq +1$) is highlighted with the yellow background. The CNO-enhanced UMP stars are displayed with larger symbols with black edges; \gdr (this work, red circles), HE~1327$-$2326 \citep[][blue]{frebel2008}, J0815$+$4729 \citep[][orange]{GonzalezHernandez2020}, and HE~0144$-$4657 \citep[][green]{placco2026umpHelmiStream}. The UMP stars in our compilation (Section \ref{subsec:lit_ump}/Appendix \ref{appendix:ump_stars}/Table \ref{tab:ump_comp}) are shown as the black dots. Those UMP stars that are also CEMP are plotted as the larger yellow dots. Stars in \texttt{JINAbase} \citep[][2021 update]{jina} with abundances similar to the LAP1-B are marked as the green squares. Stars from Milky Way UFD satellites are the black crosses (all references in Appendix \ref{appendix:ufd_stars}). The complete \texttt{JINAbase} at $-4 < \feh \leq -2$ is presented as the purple 2D histogram (Appendix \ref{appendix:refs_jinabase}).  
\label{fig:extragal_comparison}}
\end{figure*}

The increasing CEMP ($\cfe \gtrsim +1$) fraction at decreasing metallicities (specifically, iron abundance) has long been interpreted as a signature of nucleosynthesis by Population~III stars, as discussed in the review by \citet{beers2005}. Even though the excess of carbon can be explained by binary mass transfer at higher metallicities \citep[$\gtrsim$0.5\% solar;][]{yoon2016}, this mechanism likely did not have enough time to operate in the UMP regime. Recent high-redshift ($z \gtrsim 7$) \textit{JWST} observations of moderately enhanced carbon-to-oxygen ratios ($0.0 < \rm[C/O] \lesssim +0.3$) have been suggested as evidence for the CEMP phenomenon in distant galaxies \citep[][]{deugenio2024gsz12, nakajima2025pop3Gal, pollock2026highz}. However, oxygen abundances do not necessarily track iron-peak element abundances in CEMP stars, which can be appreciated from the discussion in Section \ref{sec:low-e_sn} above. To illustrate this, we compare the supposedly CEMP `ultra-faint dwarf' (UFD)\footnote{In the Local Group, UFD galaxies are defined as those with stellar masses ${\lesssim} 10^5\,\msun$ \citep[][]{simon2019}.} galaxy LAP1-B at $z \sim 6.6$ \citep[][abundances from \citealt{nakajima2025pop3Gal}]{vanzella2023lap1b} with UMP stars in our compilation, other metal-poor stars in \texttt{JINAbase}, and stars from Milky Way UFD satellites (Figure \ref{fig:extragal_comparison}).


\begin{figure*}[pt!]
\centering
\includegraphics[width=1.5\columnwidth]{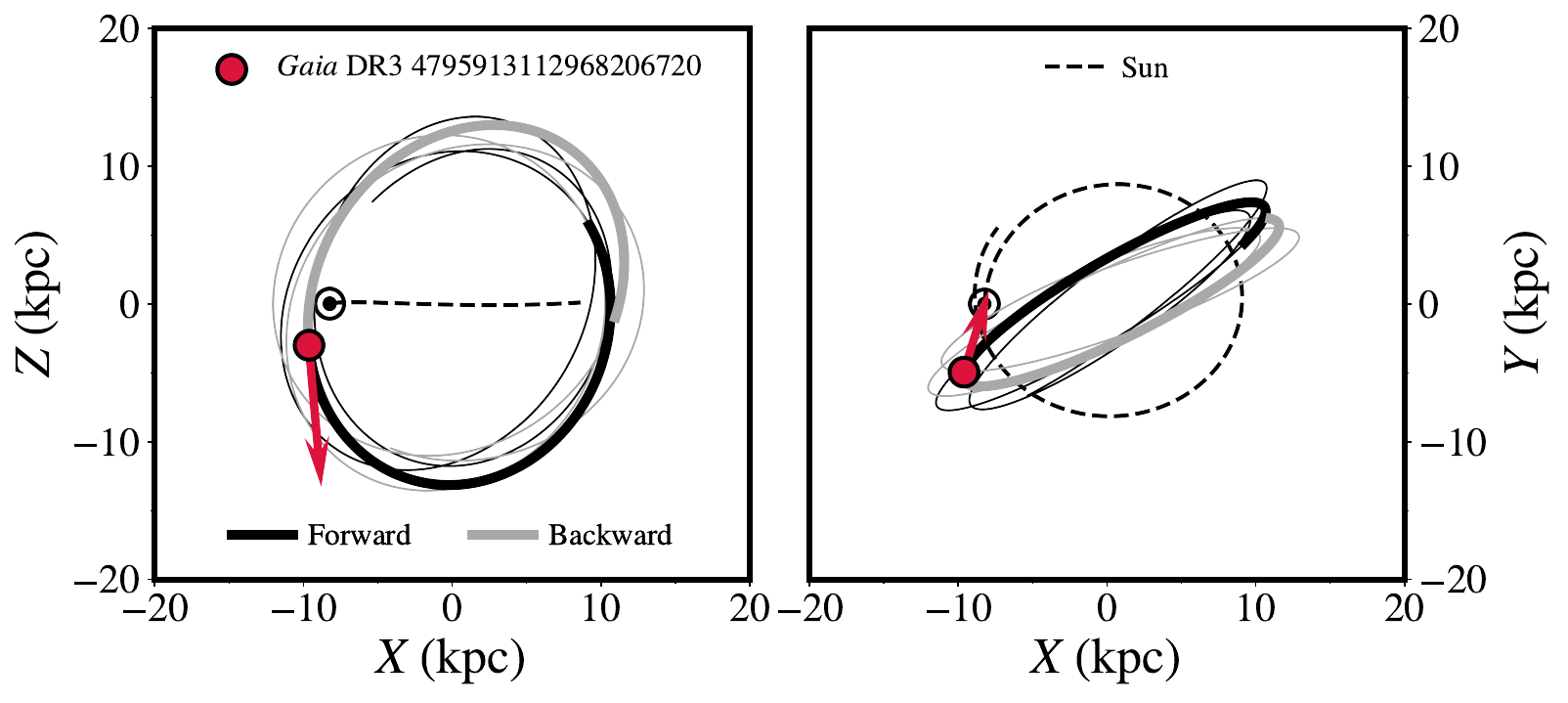}
\includegraphics[width=2.1\columnwidth]{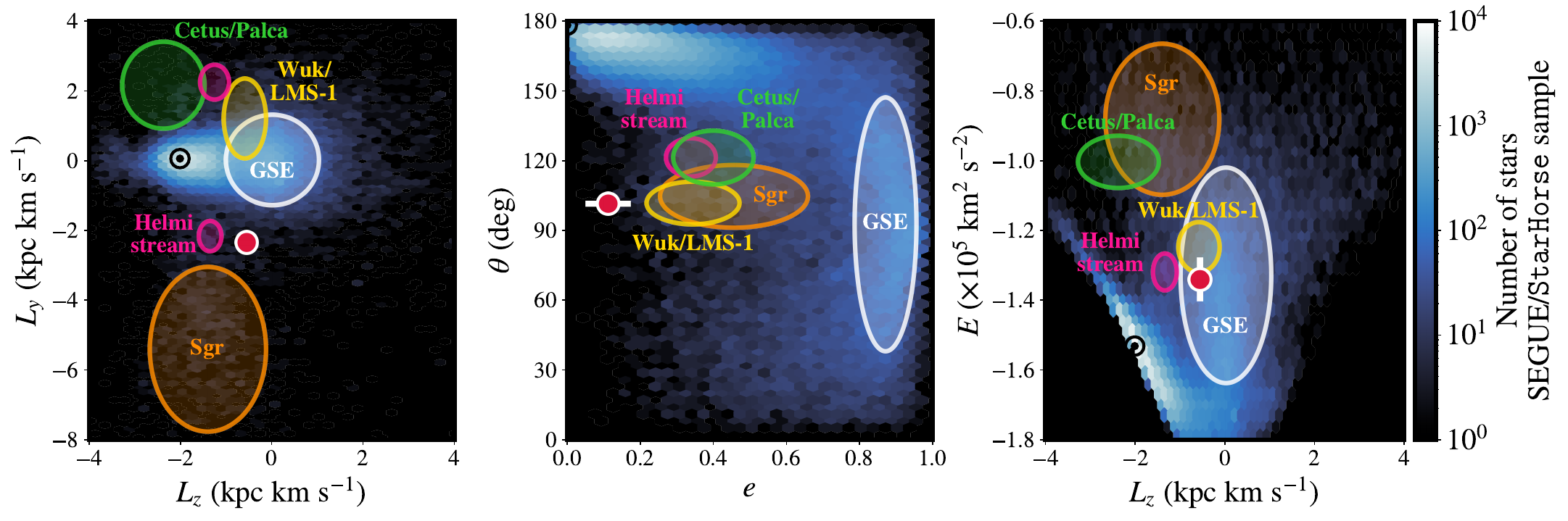}
\caption{Top panels: Orbital trajectory of \gdr (red circles) in Galactocentric Cartesian coordinates. Top left: $(X,Z)$, Milky Way edge-on view. Top right: $(X,Y)$, face on. Red arrows represent the present-day velocity vector of \gdr. Black and gray solid lines exhibit integrated trajectories forward and backward in time, respectively. Thicker and thinner lines show orbits calculated for 200\,Myr and 1\,Gyr, respectively. Dashed lines display the orbit of the Sun ($\odot$) in the same \citet[][]{mcmillan2017} model potential. Bottom panels: Various kinematic/dynamical parameter spaces. Bottom left: $(L_z, L_y)$. Bottom middle: $(e, \theta)$. Bottom right: $(L_z, E)$. The background 2D number-density histograms correspond to Milky Way metal-poor stars ($\feh \leq -0.5$) from the Sloan Extension for Galactic Understanding and Exploration (SEGUE) survey \citep[][]{yanny2009} with spectro-photometric \texttt{StarHorse} distances, as presented in \citet[][]{Limberg2023sgr}. We also highlight the \textit{loci} occupied by some of the most relevant halo substructures (see text), including \textit{Gaia}-Sausage/Enceladus (GSE, white), Sagittarius stream (Sgr, orange), Helmi~streams (pink), Wukong/LMS-1 (yellow), and Cetus/Palca (green).}
\label{fig:kindyn}
\end{figure*}

We convert the absolute abundance values of LAP1-B to the `bracket notation' assuming the \citet[][]{asplund2009} solar chemical composition (as in Section \ref{abunds} and check Footnote \ref{footnote_abund}). We immediately notice that the population of CNO-enhanced UMP stars, including \gdr, generally exhibits $\sim$10$\times$ higher [O/H] than LAP1-B (left panel in Figure \ref{fig:extragal_comparison}). On the other hand, out of the UMP stars with oxygen detection, the lowest $\rm[O/H] \sim -3$, the aforementioned star BPS~CS~30336$-$0049 \citep[][]{Lai2008_sn_yield_fits}, is $\sim$5$\times$ lower than LAP1-B. Indeed, BPS~CS~30336$-$0049 is a UMP star with one of the lowest mass-fraction metallicities ever found \citep[see figure~2 in][]{ji2026natas}. 

We then select \texttt{JINAbase} stars with similar abundances to LAP1-B ($\rm[O/H] < -2.2$ and $0.0 \leq \rm[C/O] < +0.4$; green squares in Figure \ref{fig:extragal_comparison}). We show that most of these stars are not of the CEMP kind as defined in Stellar Archaeology literature \citep[][]{beers2005, aoki2007, placco2014Carbon, Arentsen2022cemp}, i.e., the somewhat high $\rm[C/O]$ does not immediately translate to $\cfe > +1.0$ or even ${>}{+}0.7$\,dex (right panel in Figure \ref{fig:extragal_comparison}). To guarantee that a high-redshift extremely metal-poor galaxy ($-3 < \rm[O/H] < -2$\footnote{Equivalent to $5.7 < 12+\log{\rm (O/H)} < 6.7$ assuming \citet[][]{asplund2009} solar chemical composition.}) really exhibits a CEMP signature, its $\rm[C/O]$ value should be ${\gtrsim}{+}1\,{\rm dex}$, hence $\log{\rm (C/O)} \gtrsim +0.75$; a promising candidate has recently been reported by \citet{pollock2026highz}. Figure \ref{fig:all_abunds} shows the same group of stars with LAP1-B-like abundances in other chemical spaces. They generally occupy the typical \textit{locus} of very metal-poor stars ($-3 \lesssim \feh < -2$) found in the Milky Way's halo, a fraction of which do come from disrupted UFD galaxies \citep[][]{griffen2018, Brauer2019, Brauer2022}. Indeed, direct UFD chemical trends are consistent with this picture (`x' symbols in Figure \ref{fig:all_abunds}; all references in Appendix \ref{appendix:ufd_stars}). Therefore, LAP1-B abundances are largely consistent with near-field observations of low-metallicity stars both in the Milky Way's halo and its UFD satellites, but there is no evidence that it consists of a CEMP galaxy. 

\subsection{No dynamical association of \gdr with major Milky Way halo substructures}

With the kinematic parameters calculated in Section \ref{subsec:kindyn}, we test \gdr for association with known halo substructures of dwarf-galaxy origin. Figure \ref{fig:kindyn} shows the comparison between \gdr and the dynamical parameter space occupied by the debris from the most massive Milky Way accretion events, including Sagittarius dwarf spheroidal galaxy (Sgr) and its tidal stream \citep[e.g.,][]{Ibata1994, Majewski2003}, \textit{Gaia}-Sausage/Enceladus \citep[GSE;][]{belokurov2018, Haywood2018, helmi2018}, and others (see below). The selection criteria is adopted from \citet[][]{Limberg2023sgr} for Sgr and the combined \citet[][]{Feuillet2020}, \citet[][]{naidu2020}, and \citet[][]{Limberg2022gse} for GSE. The orbital energy of \gdr ($E \sim -1.3 \times 10^5\,{\rm km}^2\,{\rm s}^{-2}$ in the \citealt{mcmillan2017} model potential) is too low to be consistent with Sgr. At the same time, the almost-circular orbit of \gdr ($e \sim 0.1$) also makes it incompatible with GSE, which is typically characterized by $e \gtrsim 0.8$ \citep[e.g.,][]{naidu2020}.

In the bottom panels of Figure \ref{fig:kindyn}, we also compare \gdr's kinematics to other disrupted dwarf galaxies \citep[e.g.,][]{Malhan2021lms1, Malhan2022atlas} following the appendix in \citet{Limberg2023sgr}. \gdr has $L_y = -2351\,\angmom$ with an opposite sign from both Cetus/Palca \citep{Newberg2009cetus, Shipp2018, Yuan2019cetus, Thomas&Battaglia2022cetus} and Wukong/LMS-1 \citep{naidu2020, Yuan2020lms1, Limberg2024wuk} stellar streams. \gdr's $L_z = -555\,\angmom$ and specific energy are also not large enough to be consistent with Cetus/Palca. The orbital inclination of \gdr is too extreme $\theta \sim 100\,{\rm deg}$ to be compatible with either Cetus/Palca or the Helmi~streams \citep{helmi1999,myeongStreamsAndClumps, myeongShards, koppelmanHelmi, Limberg2021hstr}. Lastly, the orbital eccentricity of \gdr is also too small, i.e., the orbit is too circular, in contrast to any of the aforementioned substructures.

This lack of association to any of the known relatively bright disrupted dwarf galaxies (${\rm progenitor \ stellar \ masses } \gtrsim 10^5\,\msun$) in the Milky Way's halo is consistent with the UFD origin for \gdr, following the direct detection of a CEMP UMP star in a UFD galaxy by \citet[][]{chiti2026pic2}. Given the low stellar masses of UFD galaxies, it is extremely challenging to identify their tidal remnants \citep[][]{Yuan2020dtgs, Gudin2021, Limberg2021dtgs}. Although the Galactic halo is expected to host hundreds of destroyed UFDs, current techniques are unable to confidently distinguish them as coherent dynamical groups against the overwhelming smooth halo fore/background \citep[][]{Brauer2022}.

\section{Conclusions} \label{conclusions}

We have reported the independent recovery and chemo-dynamical analysis of \gdr, a CEMP UMP star ($\feh = -4.06$ and $\cfe = +1.54$) in the \textit{Gaia}~XP catalog. 
\gdr had been previously found to be a UMP star by \citet{yong2021_SMSSmetalpoor}. Here, we have measured additional abundances, derived a distance estimate, and performed a comprehensive dynamical analysis, thus contextualizing our findings with stellar populations in the Milky Way's halo and its UFD satellites. We have detected both nitrogen ($\rm[N/Fe] = 2.64$) and oxygen ($\rm[O/Fe] = 2.62$) in \gdr's atmosphere, making this object one of only 5 UMP stars ($\feh < -4$) with an [O/Fe] measurement and one of just 4 CNO-enhanced UMP stars. We have found the abundances of other light elements in \gdr to also be well above the solar level, including Na, Mg, Al, and Si.

We have fitted the full abundance pattern of \gdr with Population~III supernova yield models and found that it is best reproduced by a 21.5\,\msun progenitor and a low-energy explosion ($E_{\rm SN} = 0.3 \times 10^{51}\,{\rm erg}$). Our results are consistent with other CEMP and CNO-enhanced UMP stars, corroborating the faint supernovae with a mixing-and-fallback mechanism \citep[][]{Umeda2003, iwamoto2005faintSN} as an important nucleosynthesis pathway in the early Universe. Since the excess of carbon in low-metallicity environments can be accompanied by overproduction of oxygen, as well as other light elements, in low-energy supernovae, we have shown that the [C/O] ratio is a poor diagnostic of the CEMP phenomenon in high-redshift galaxies now being observed with \textit{JWST}. Finally, orbital dynamics of \gdr suggest that this UMP star likely formed in a low-mass dwarf-galaxy environment, presumably in a UFD system, and later accreted onto the Milky Way.

\begin{acknowledgments}

G. L. is indebted to all those involved with the multi-institutional \textit{MilkyWayBR} Group for the weekly discussions. G.L. acknowledges support from KICP/UChicago through a KICP Postdoctoral Fellowship. G.L. also thanks Erik Solhaug for conversations about gas-phase abundances and Madeleine McKenzie for conversations about stellar parameters. The work of V.M.P. is supported by NOIRLab, which is managed by the Association of Universities for Research in Astronomy (AURA) under a cooperative agreement with the U.S. National Science Foundation. F.A. acknowledges financial support from MCIN/AEI/10.13039/501100011033 through a RYC2021-031638-I grant co-funded by the European Union NextGenerationEU/PRTR. This work was partially supported by the Spanish MICIN/AEI/10.13039/501100011033 and by ‘ERDF A way of making Europe’ by the European Union through grant PID2024-157964OB-C21, and the Institute of Cosmos Sciences University of Barcelona (ICCUB, Unidad de Excelencia Mar\'{\i}a de Maeztu) through grant CEX2024-001451-M and the project 2021-SGR-00679 GRC of the Agència de Gestió d'Ajuts Universitaris i de Recerca (Generalitat de Catalunya). A.dG. acknowledges support from a Clay Fellowship awarded by the Smithsonian Astrophysical Observatory.

This work has used data from the European Space Agency (ESA) mission {\it Gaia} (\url{https://www.cosmos.esa.int/gaia}), processed by the {\it Gaia} Data Processing and Analysis Consortium (DPAC, \url{https://www.cosmos.esa.int/web/gaia/dpac/consortium}). Funding for the DPAC has been provided by national institutions, in particular the institutions participating in the {\it Gaia} Multilateral Agreement.

This publication makes use of data products from the Two Micron All Sky Survey, which is a joint project of the University of Massachusetts and the Infrared Processing and Analysis Center/California Institute of Technology, funded by the National Aeronautics and Space Administration and the National Science Foundation.

This research has made use of the Astrophysics Data System, funded by NASA under Cooperative Agreement 80NSSC21M0056.



\end{acknowledgments}

\setlength{\bibsep}{0pt}
\bibliographystyle{aasjournalv71}
\footnotesize

\bibliography{bibliography.bib}{}

\clearpage
\appendix

\section{Literature UMP stars}
\label{appendix:ump_stars}

The complete list of UMP stars constructed in Section \ref{subsec:lit_ump} is in Table \ref{tab:ump_comp}, including literature name, \textit{Gaia} DR3 \texttt{source\_id}, reported \feh, and adopted reference.

\section{Evolutionary carbon-abundance corrections with \texttt{carbcor}}
\label{carbcor}

As described in Section \ref{subsec:lit_ump}, we automate the task of deriving carbon-abundance corrections for large samples of stars by developing the \texttt{carbcor} Python package \citep{carbcor}. Basically, \texttt{carbcor} takes input \logg, [Fe/H], and measured [C/Fe] from a list of stars and fetches the corrections from the online tool provided by \citet[][]{placco2014Carbon}. As of now (v1.0.2), since \texttt{carbcor} only parses one star at a time, it can only obtain carbon corrections for $\sim$2500 stars in 10\,min. Even so, this new method has the advantage of making our results that require [C/Fe] immediately reproducible. We also make \texttt{carbcor} installable via the Python Package Index\footnote{\url{https://pypi.org/project/carbcor/}.} with a simple command line:
\begin{lstlisting}[language=Python]
pip install carbcor.
\end{lstlisting}
To retrieve [C/Fe] correction for \gdr, for example, the user can enter the command line 
\begin{lstlisting}[language=bash]
carbcor query 2.02 -4.06 1.53
\end{lstlisting}
in a terminal, where $\logg = 2.02$, $\feh = -4.06$, and $\cfe = 1.53$; the output is $\mathtt{cfe\_corrected} = 1.54$ (Section \ref{abunds}). It also provides the option to input larger batches of stars in \texttt{pandas} DataFrame format \citep[][]{pandas}. The \texttt{carbcor} package makes carbon evolutionary corrections a lot more accessible and easy to incorporate in modern analysis frameworks for future studies of metal-poor RGB stars.

\section{Complete atomic data and abundances for \gdr}
\label{appendix:atomic_data}

Table \ref{eqwl} shows the line-by-line absolute abundances derived for \gdr in Section \ref{abunds}, as well as the atomic and molecular data (excitation potentials and transition probabilities) and NLTE corrections. Abundances determined via spectral synthesis are marked as ``\texttt{syn}'' on the equivalent width (EW) column.

\section{Systematic abundance uncertainties for \gdr}
\label{appendix:sys_uncs}

Table \ref{sys} depicts systematic abundance uncertainties for \gdr due to stellar-parameter variations, as described in Section \ref{abunds}. The values in ``$\sigma$'' column are the standard error of the mean, and the final uncertainty $\sigma_{\rm tot}$ consists of the quadrature sum of the individual uncertainty estimates.

\section{Literature references for stars in \texttt{JINAbase}}
\label{appendix:refs_jinabase}

Here, we list the references for the individual works that compose the updated \texttt{JINAbase} with stars at $\feh \leq -2$, in alphabetical order and sorted by the year of publication for authors with multiple contributions; \citet[][]{Afsar2016jina, afsar2018jina, allen2012jina, Andrievsky2009jina, Andrievsky2010jina}, \citet[][]{Aoki2002jinaA, aoki2002jinaC, aoki2002jinaD, aoki2002jinaB, aoki2005jina, aoki2006jina, aoki2007, aoki2008jina, aoki2010jina, aoki2012jina, aoki2013jina, aoki2014jina, aoki2017jina, arnone2005jina, Bandyopadhyay2020, barbuy2005jina, barklem2005, Battistini2015jina, behara2010jina, bensby2011jina, Bensby2014, bergemann2010jinaA, bergemann2010jinaB, bonifacio2009jina, bonifacio2012jina, Boesgaard2011jina, burris2000jina, caffau2011jina, caffau2013jina, caffau2019jina, Caliskan2014jina, cain2018jina, Carretta2002jina, Casey2014orphan, casey2017jina, Casey2015, Cayrel2004, cohen2003jina, cohen2004jina, cohen2006jina, cohen2008, Cohen2013, cowan2005jina, cowan2020jina, cui2013jina, duong2019jina, ezzedine2020, ForSneden2010, frebel2007jinaB, frebel2007jinaA, Fulbright2000, Gallagher2010jina, GarciaPerez2009jina, Goswami2016jina, gull2018jina, hansen2011jina, hansen2012jina, Hansen2019, hansen2020jina, hansen2014, hansen2018, hayek2009jina, Hill2002, hill2017jina, hollek2011jina, hollek2015jina, holmbeck2018, holmbeck2020, honda2004jina, honda2006jina, Honda2007jina, Honda2011jina, Hosford2009jina, howes2015jina, howes2016jina, ishigaki2010jina, ishigaki2012jina, ishigaki2013, ivans2003jina, ivans2005jina, ivans2006jina, Jacobson2015metalpoor, Johnson2002jinaA, Johnson2002jinaB, Johnson2004jina, Jonsell2005jina, Jonsell2006jina, kennedy2014, koch2016bulge, Korotin2018, Lai2007jina, Lai2008_sn_yield_fits, Lai2009jina, Li2013fluorine, LiHaining2015lamost, li2015early, HainingLi2018, Lucatello2003, mardini2019_II, Mardini2019, Mardini2020, mashonkina2010hes, mashonkina2014, masseron2006, masseron2012, mcwilliam1995, melendez2010spitePlateau, navarrete2015ocen, norris1997_CS22957_027, norris1997_empCEMP, norris1997_Li, norris2000_CS22876_032, norris2001, norris2002, omalley2017diff, OuX2020vanadium, peterson2011_molybdenum, peterson2013_molybdenum, Placco2013, placco2014High-Res, placco2015seguefollowup, placco2015hst, Placco2016B, Placco2017, Placco2020, preston2000, preston2001, preston2006, Rasmussen2020, Reggiani2017, Reggiani2020, ren2012hes, rich2009_BeO, roederer2008jina, roederer2009jina, Roederer2010, roederer2012hst, roederer2012tellurium, roederer2014subgiant, roederer2014Phosphorus, Roederer2014ncap, roederer2014, roederer2014heaveElems, Roederer2016CEMP, roederer2018hd222925, roederer2020_Pb, roederer2018_copper_zinc, Roederer2019sylgr, Ruchti2011jina, SeanRyan1991, Ryan1996, Saito2009zinc, sakari2018, schuler2007fluorine, SiqueiraMello2012_hd140283, siqueira2014, siqueiramello2015, Sivarani2004_CS29497_030, Sivarani2006, Smiljanic2009beryllium, sneden2003, sneden2009, sneden2016ironpeak, spite2000, spite2011, spite2012, spite2013, spite2014, spite2019, SusmithaRani2016, takeda2011jina, tan2009beryllium, wes2000, yong2013full, zhang2009haloStars}.

\section{Literature references for stars in UFD galaxies}
\label{appendix:ufd_stars}

References for literature UFD galaxy stars, in alphabetic order, shown in Figure \ref{fig:all_abunds}; Bo\"otes~I \citep[][]{Feltzing2009bootes1, norris2010bootes1, norris2010bootes1_EMP, lai2011bootes1, gilmore2013bootes1, ishigaki2014bootes1, frebel2016bootes1, waller2023bootes1}, Bo\"otes~II \citep{ji2016bootes1}, Carina~II \citep{Ji2020MagLiteS}, Carina~III \citep{Ji2020MagLiteS}, Centaurus~I \citep[][]{heiger2025_Cen1Eri4}, Cetus~II \citep[][]{webber2023cetus2}, Coma~Berenices \citep[][]{frebel2010_UMa2_ComBer, waller2023bootes1}, Canes~Venatici~II \citep[][]{francois2016ufds}, Eridanus~IV \citep[][]{heiger2025_Cen1Eri4}, Grus~I \citep[][]{ji2019_Gru1Tri2}, Grus~II \citep[][]{hansen2020grus2}, Hercules \citep[][]{koch2008hercules, koch2013hercules}, Horologium~I \citep{nagasawa2018horologium1}, Leo IV \citep[][]{simon2010leo4, francois2016ufds}, Pisces~II \citep[][]{spite2018pisces2}, Reticulum~II \citep[][]{Ji2016b, roederer2016}, Segue~1 \citep{frebel2014}, Segue~2 \citep[][]{roederer2014segue2}, Triangulum~II \citep[][]{ji2019_Gru1Tri2}, Tucana~II \citep[][]{chiti2018, chiti2023tucana2}, Tucana~V \citep[][]{hansen2024tucana5}, Ursa~Major~I \citep[][]{waller2023bootes1}, Ursa~Major~II \citep[][]{frebel2010_UMa2_ComBer}.

\renewcommand{\arraystretch}{1.0}
\setlength{\tabcolsep}{1.0em}
\begin{deluxetable*}{@{}lrrl@{}}[!ht]
\tabletypesize{\small}
\tabletypesize{\footnotesize}
\tablewidth{0pc}
\tablecaption{Curated list of literature UMP stars \label{tab:ump_comp}}
\tablehead{
\colhead{Literature name}&
\colhead{\textit{Gaia} DR3 \texttt{source\_id}}&
\colhead{\feh}&
\colhead{Reference}
}
\startdata
BPS CS 30324$-$0063          & 2367173119271988480 & $-$4.05            & \citet[][]{Placco2016B} \\
J0023$+$0307                 & 2548541852945056896 & $<-6.30$  & \citet[][]{Frebel2019} \\
CD$-$38 245                  & 5000753194373767424 & $-$4.20             & \citet[][]{Cayrel2004}   \\
AS0039                     & 5003213763958906496 & $-$4.11            & \citet[][]{Skuladottir2021UMPsculptor}   \\
HE 0107$-$5240                & 4927204800008334464 & $-$5.44            & \citet[][]{Christlieb2004he0107-5240}      \\
HE 0134$-$1519                & 2453397508316944128 & $-$3.98            & \citet[][]{hansen2014}      \\
SDSS J014036.21+234458.1   & 290930261314166528  & $-$4.00               & \citet[][]{bonifacio2018topos}      \\
HE 0144$-$4657                & 4954178397218620800 & $-$4.11            & \citet[][]{placco2026umpHelmiStream}    \\
SMSS J022423.27$-$573705.1    & 4739093513140451200 & $-$3.97            & \citet[][]{Jacobson2015metalpoor}    \\
BD$+$44 493                  & 341511064663637376  & $-$3.96            & \citet[][]{placco2024}      \\
HE 0233$-$0343                & 2495327693479473408 & $-$4.68            & \citet[][]{hansen2014}   \\
BPS CS 22963$-$0004            & 5184426749232471808 & $-$4.09            & \citet{roederer2014}      \\
SMSS J031300.36$-$670839.3   & 4671418400651900544 & $<-7.30$  & \citet[][]{keller2014}    \\
G 77$-$61                     & 3265069670684495744 & $-$4.08            & \citet[][]{plez2005ump}        \\
J0422+1808                 & 47471831142473216   & $-$4.06            & \citet[][]{Li2022lamost}    \\
MAGIC J043330.99−554839.2  & 4775927771146196864 & $-$4.12            & \citet[][]{placco2025magic}          \\
\textbf{GDR3 4795913112968206720}   & \textbf{4795913112968206720} & $-$4.06            & \textbf{This work}        \\
LMC-119                    & 4756487787099363968 & $-$4.13            & \citet[][]{chiti2024lmc}        \\
HE 0557$-$4840                & 4794791782906532608 & $-$4.73            & \citet[][]{norris2007}     \\
MAGIC PicII-503           & 5480105831331104384 & $<-4.63$ & \citet[][]{chiti2026pic2} \\
GDR3 5262850721755411072   & 5262850721755411072 & $-$4.82            & \citet[][]{limberg2025ump_gaiaxp}       \\
J0815$+$4729                 & 931227322991970560  & $-$5.49            & \citet[][]{GonzalezHernandez2020} \\
SDSS J092912.32$+$023817.0    & 3844818546870217728 & $-$4.97            & \citet[][]{caffau2016}      \\
HE 1012$-$1540                & 3751852536639575808 & $-$4.17            & \citet{roederer2014}     \\
SDSS J102915.14$+$172927.9   & 3890626773968983296 & $-$4.73            & \citet[][]{caffau2024}        \\
SDSS J103402.71$+$070116.6   & 3862721340654330112 & $-$4.01            & \citet[][]{bonifacio2018topos}    \\
SDSS J103556.11$+$064143.9   & 3862507691800855040 & $<-5.20$  & \citet[][]{bonifacio2018topos}      \\
SDSS J120441.38$+$120111.5   & 3919025342543602176 & $-$4.34            & \citet[][]{placco2015seguefollowup}  \\
SDSS J124719.46$-$034152.4   & 3681866216349964288 & $-$4.01            & \citet[][]{bonifacio2018topos}    \\
SDSS J131326.89$-$001941.4    & 3687441358777986688 & $-$5.00               & \citet[][]{Frebel2015ApJ}    \\
HE 1310$-$0536                & 3635533208672382592 & $-$4.15            & \citet[][]{hansen2014}  \\
HE 1327$-$2326     & 6194815228636688768 & $-$5.71            & \citet[][]{frebel2008}     \\
HE 1424$-$0241                & 3643332182086977792 & $-$4.14            & \citet[][]{Cohen2013}        \\
SDSS J144256.37$-$001542.7   & 3651420563283262208 & $-$4.37            & \citet[][]{bonifacio2018topos}     \\
Pristine 221.8781$+$9.7844  & 1174522686140620672 & $-$4.64            & \citet[][]{starkenburg2018ump}  \\
SMSS J160540.18$-$144323.1   & 6262975053962303744 & $-$6.21            & \citet[][]{nordlander2019smss}    \\
SDSS J174259.67$+$253135.8    & 4581822389265279232 & $-$4.77            & \citet[][]{Bonifacio2015}     \\
2MASS J18082002$-$5104378   & 6702907209758894848 & $-$4.07            & \citet[][]{Melendez2016}    \\
BPS CS 22891$-$0200          & 6445220927325014016 & $-$4.06            & \citet{roederer2014}     \\
BPS CS 22885$-$0096   & 6692925538259931136 & $-$4.41            & \citet{roederer2014}      \\
BPS CS 22950$-$0046   & 6876806419780834048 & $-$4.12            & \citet{roederer2014}     \\
BPS CS 30336$-$0049    & 6795730493933072128 & $-$4.04            & \citet[][]{Lai2008_sn_yield_fits}      \\
J2050$−$6613                 & 6425379106130933376 & $-$4.05            & \citet[][]{mardini2024diskUMP}     \\
SPLUS J210428.01$-$004934.2   & 2689845933385992064 & $-$4.19            & \citet[][]{placco2021ump}      \\
HE 2139$-$5432                & 6461736966363075200 & $-$4.02            & \citet[][]{yong2013full} \\
J2217$+$2104                 & 1778804140643594240 & $-$4.12            & \citet[][]{Li2022lamost}         \\
HE 2239$-$5019                & 6513870718215626112 & $-$4.15            & \citet[][]{hansen2014}    \\
SDSS J230959.55$+$230803.0   & 2838938855416252160 & $-$3.96            & \citet[][]{matsuno2017seguefollowup}  \\
BPS CS 22949-0037          & 2634585342263017984 & $-$3.99            & \citet[][]{Cayrel2004} \\
HE 2323-0256                & 2634585342263017984 & $-$3.98            & \citet[][]{Cohen2013}
\enddata
\end{deluxetable*}

\clearpage
\startlongtable
\setlength{\tabcolsep}{0.5em}
\begin{deluxetable*}{lrrrrrr}
\tabletypesize{\tiny}
\tabletypesize{\footnotesize}
\tablewidth{0pc}
\tablecaption{\label{eqwl} Atomic Data and Derived Abundances}
\tablehead{
\colhead{Ion}&
\colhead{$\lambda$}&
\colhead{$\chi$} &
\colhead{$\log\,gf$}&
\colhead{EW}&
\colhead{$\log\epsilon$\,(X)}&
\colhead{$\Delta$NLTE}\\
\colhead{}&
\colhead{({\AA})}&
\colhead{(eV)} &
\colhead{}&
\colhead{(m{\AA})}&
\colhead{}&
\colhead{}}
\startdata
  C (CH)     &  4307.00 & \nodata & \nodata &     \texttt{syn} &    5.880 & \nodata \\ 
  C (CH)     &  4323.00 & \nodata & \nodata &     \texttt{syn} &    5.920 & \nodata \\ 
  N (NH)     &  3359.00 & \nodata & \nodata &     \texttt{syn} &    6.410 & \nodata \\ 
  N (CN)     &  3883.00 & \nodata & \nodata &     \texttt{syn} &    6.410 & \nodata \\ 
 \ion{O}{1}  &  7771.94 &    9.15 &    0.35 &     \texttt{syn} &    7.290 & \nodata \\ 
 \ion{O}{1}  &  7774.17 &    9.15 &    0.22 &     \texttt{syn} &    7.290 & \nodata \\ 
 \ion{O}{1}  &  7775.39 &    9.15 &   -0.02 &     \texttt{syn} &    7.170 & \nodata \\ 
 \ion{Na}{1} &  5889.95 &    0.00 &    0.11 &  127.41 &    3.735 & $-$0.52 \\
 \ion{Na}{1} &  5895.92 &    0.00 & $-$0.19 &  117.43 &    3.828 & $-$0.46 \\
 \ion{Mg}{1} &  3829.35 &    2.71 & $-$0.23 &  130.36 &    4.959 &    0.28 \\
 \ion{Mg}{1} &  3832.30 &    2.71 &    0.25 &  158.37 &    4.876 &    0.24 \\
 \ion{Mg}{1} &  3838.29 &    2.72 &    0.47 &  161.68 &    4.701 &    0.22 \\
 \ion{Mg}{1} &  4702.99 &    4.33 & $-$0.44 &   34.88 &    4.863 &    0.20 \\
 \ion{Mg}{1} &  5172.68 &    2.71 & $-$0.36 &  137.28 &    4.917 &    0.27 \\
 \ion{Mg}{1} &  5183.60 &    2.72 & $-$0.17 &  148.48 &    4.904 &    0.26 \\
 \ion{Mg}{1} &  5528.40 &    4.35 & $-$0.55 &   30.40 &    4.890 &    0.21 \\
 \ion{Mg}{1} &  8806.76 &    4.35 & $-$0.14 &   64.45 &    4.917 &    0.17 \\
 \ion{Al}{1} &  3944.00 &    0.00 & $-$0.64 &     \texttt{syn} &    2.170 & \nodata \\ 
 \ion{Al}{1} &  3961.52 &    0.01 & $-$0.33 &     \texttt{syn} &    2.250 &    0.80 \\ 
 \ion{Si}{1} &  3905.52 &    1.91 & $-$1.04 &     \texttt{syn} &    4.240 &    0.11 \\ 
 \ion{Ca}{1} &  4226.74 &    0.00 &    0.24 &     \texttt{syn} &    2.440 &    0.35 \\ 
 \ion{Sc}{2} &  3576.34 &    0.01 &    0.01 &   34.82 & $-$0.832 & \nodata \\
 \ion{Sc}{2} &  3630.74 &    0.01 &    0.22 &   42.34 & $-$0.887 & \nodata \\
 \ion{Sc}{2} &  4246.82 &    0.32 &    0.24 &   39.99 & $-$0.773 & \nodata \\
 \ion{Sc}{2} &  4320.73 &    0.60 & $-$0.25 &    9.96 & $-$0.776 & \nodata \\
 \ion{Ti}{2} &  3759.29 &    0.61 &    0.28 &   89.91 &    1.324 &    0.19 \\
 \ion{Ti}{2} &  3761.32 &    0.57 &    0.18 &   84.68 &    1.192 &    0.19 \\
 \ion{Ti}{2} &  3913.46 &    1.12 & $-$0.36 &   48.69 &    1.323 &    0.09 \\
 \ion{Ti}{2} &  4395.03 &    1.08 & $-$0.54 &   36.26 &    1.147 &    0.07 \\
 \ion{Ti}{2} &  4399.77 &    1.24 & $-$1.20 &   10.80 &    1.295 &    0.10 \\
 \ion{Ti}{2} &  4417.71 &    1.17 & $-$1.19 &    9.42 &    1.141 &    0.10 \\
 \ion{Ti}{2} &  4443.80 &    1.08 & $-$0.71 &   30.73 &    1.200 &    0.01 \\
 \ion{Ti}{2} &  4501.27 &    1.12 & $-$0.77 &   22.03 &    1.098 &    0.11 \\
 \ion{Ti}{2} &  4533.96 &    1.24 & $-$0.53 &   26.43 &    1.095 &    0.12 \\
 \ion{Ti}{2} &  4571.97 &    1.57 & $-$0.31 &   20.36 &    1.083 &    0.03 \\
 \ion{Cr}{1} &  4274.80 &    0.00 & $-$0.22 &   20.44 &    1.192 &    0.91 \\
 \ion{Cr}{1} &  4289.72 &    0.00 & $-$0.37 &   14.51 &    1.157 &    0.91 \\
 \ion{Mn}{1} &  4030.75 &    0.00 & $-$0.50 &     \texttt{syn} &    0.550 &    1.00 \\ 
 \ion{Mn}{1} &  4033.06 &    0.00 & $-$0.65 &     \texttt{syn} &    0.610 &    1.00 \\ 
 \ion{Fe}{1} &  3618.77 &    0.99 & $-$0.00 &   81.44 &    3.464 &    0.33 \\
 \ion{Fe}{1} &  3727.62 &    0.96 & $-$0.61 &   72.94 &    3.544 &    0.43 \\
 \ion{Fe}{1} &  3758.23 &    0.96 & $-$0.01 &   88.36 &    3.430 &    0.43 \\
 \ion{Fe}{1} &  3763.79 &    0.99 & $-$0.22 &   83.65 &    3.516 &    0.45 \\
 \ion{Fe}{1} &  3787.88 &    1.01 & $-$0.84 &   56.44 &    3.346 &    0.43 \\
 \ion{Fe}{1} &  3815.84 &    1.48 &    0.24 &   78.80 &    3.410 &    0.45 \\
 \ion{Fe}{1} &  3820.43 &    0.86 &    0.16 &  104.34 &    3.573 &    0.38 \\
 \ion{Fe}{1} &  3825.88 &    0.91 & $-$0.02 &   93.93 &    3.511 &    0.41 \\
 \ion{Fe}{1} &  3827.82 &    1.56 &    0.09 &   70.78 &    3.397 &    0.44 \\
 \ion{Fe}{1} &  3840.44 &    0.99 & $-$0.50 &   70.72 &    3.355 &    0.43 \\
 \ion{Fe}{1} &  3841.05 &    1.61 & $-$0.04 &   61.22 &    3.313 &    0.43 \\
 \ion{Fe}{1} &  3849.97 &    1.01 & $-$0.86 &   63.03 &    3.517 &    0.30 \\
 \ion{Fe}{1} &  3865.52 &    1.01 & $-$0.95 &   50.30 &    3.297 &    0.42 \\
 \ion{Fe}{1} &  3878.02 &    0.96 & $-$0.90 &   56.43 &    3.329 &    0.41 \\
 \ion{Fe}{1} &  3902.95 &    1.56 & $-$0.44 &   49.18 &    3.361 &    0.41 \\
 \ion{Fe}{1} &  4005.24 &    1.56 & $-$0.58 &   42.13 &    3.335 &    0.40 \\
 \ion{Fe}{1} &  4045.81 &    1.49 &    0.28 &   86.76 &    3.558 &    0.43 \\
 \ion{Fe}{1} &  4063.59 &    1.56 &    0.06 &   72.93 &    3.432 &    0.43 \\
 \ion{Fe}{1} &  4071.74 &    1.61 & $-$0.01 &   68.18 &    3.420 &    0.42 \\
 \ion{Fe}{1} &  4143.87 &    1.56 & $-$0.51 &   49.50 &    3.404 &    0.41 \\
 \ion{Fe}{1} &  4202.03 &    1.49 & $-$0.69 &   54.32 &    3.603 &    0.40 \\
 \ion{Fe}{1} &  4250.12 &    2.47 & $-$0.38 &   15.68 &    3.467 &    0.39 \\
 \ion{Fe}{1} &  4250.79 &    1.56 & $-$0.71 &   40.02 &    3.393 &    0.39 \\
 \ion{Fe}{1} &  4260.47 &    2.40 &    0.08 &   26.86 &    3.238 &    0.38 \\
 \ion{Fe}{1} &  4271.15 &    2.45 & $-$0.34 &   16.16 &    3.420 &    0.41 \\
 \ion{Fe}{1} &  4271.76 &    1.49 & $-$0.17 &   67.17 &    3.380 &    0.41 \\
 \ion{Fe}{1} &  4325.76 &    1.61 &    0.01 &   73.99 &    3.508 &    0.42 \\
 \ion{Fe}{1} &  4375.93 &    0.00 & $-$3.00 &   17.84 &    3.427 &    0.40 \\
 \ion{Fe}{1} &  4404.75 &    1.56 & $-$0.15 &   66.63 &    3.400 &    0.47 \\
 \ion{Fe}{1} &  4415.12 &    1.61 & $-$0.62 &   44.45 &    3.430 &    0.43 \\
 \ion{Fe}{1} &  4427.31 &    0.05 & $-$2.92 &   24.55 &    3.578 &    0.40 \\
 \ion{Fe}{1} &  5371.49 &    0.96 & $-$1.64 &   41.50 &    3.596 &    0.45 \\
 \ion{Fe}{1} &  5397.13 &    0.92 & $-$1.98 &   25.92 &    3.564 &    0.45 \\
 \ion{Fe}{1} &  5405.77 &    0.99 & $-$1.85 &   25.98 &    3.513 &    0.45 \\
 \ion{Fe}{2} &  4923.92 &    2.89 & $-$1.26 &   22.59 &    3.430 &    0.02 \\
 \ion{Fe}{2} &  5018.43 &    2.89 & $-$1.10 &   28.29 &    3.402 & \nodata \\
 \ion{Co}{1} &  3845.47 &    0.92 &    0.06 &   16.23 &    1.245 & \nodata \\
 \ion{Co}{1} &  3995.31 &    0.92 & $-$0.18 &   13.76 &    1.383 & \nodata \\
 \ion{Ni}{1} &  3524.54 &    0.03 &    0.01 &   99.69 &    2.742 & \nodata \\
 \ion{Ni}{1} &  3566.37 &    0.42 & $-$0.24 &   76.64 &    2.644 & \nodata \\
 \ion{Ni}{1} &  3597.70 &    0.21 & $-$1.10 &   64.75 &    2.832 & \nodata \\
 \ion{Ni}{1} &  3783.53 &    0.42 & $-$1.40 &   40.08 &    2.586 & \nodata \\
 \ion{Sr}{1} &  4077.71 &    0.00 &    0.15 &     \texttt{syn} & $-$1.850 & \nodata \\ 
 \ion{Sr}{1} &  4215.52 &    0.00 & $-$0.17 &     \texttt{syn} & $-$1.810 & \nodata \\ 
 \ion{Ba}{1} &  4554.03 &    0.00 &    0.16 &     \texttt{syn} & $-$2.420 & \nodata \\ 
\enddata
\end{deluxetable*}



\begin{deluxetable*}{@{}lrrccc@{}}[!ht]
\tabletypesize{\small}
\tabletypesize{\footnotesize}
\tablewidth{0pc}
\tablecaption{Systematic Abundance Uncertainties \label{sys}}
\tablehead{
\colhead{Ion}&
\colhead{$\Delta$\teff}&
\colhead{$\Delta$\logg}&
\colhead{$\Delta\xi$}&
\colhead{$\sigma$}&
\colhead{$\sigma_{\rm tot}$\vv}\\
\colhead{}&
\colhead{$+$60\,K}&
\colhead{$+$0.20 dex}&
\colhead{$+$0.30 \kmsec}&
\colhead{}&
\colhead{}}
\startdata
C           &    0.15 & $-$0.07 &    0.01 &    0.10 &    0.19 \\
N           &    0.16 & $-$0.10 & $-$0.03 &    0.20 &    0.28 \\
\ion{O}{1}  & $-$0.06 &    0.09 &    0.00 &    0.15 &    0.18 \\
\ion{Na}{1} &    0.06 & $-$0.02 & $-$0.16 &    0.05 &    0.18 \\
\ion{Mg}{1} &    0.05 & $-$0.05 & $-$0.08 &    0.07 &    0.13 \\
\ion{Al}{1} &    0.06 & $-$0.01 & $-$0.07 &    0.15 &    0.18 \\
\ion{Si}{1} &    0.07 & $-$0.03 & $-$0.16 &    0.15 &    0.23 \\
\ion{Ca}{1} &    0.07 & $-$0.01 & $-$0.17 &    0.10 &    0.21 \\
\ion{Sc}{2} &    0.06 &    0.05 & $-$0.03 &    0.05 &    0.10 \\
\ion{Ti}{2} &    0.04 &    0.06 & $-$0.07 &    0.09 &    0.13 \\
\ion{Cr}{1} &    0.08 & $-$0.01 & $-$0.01 &    0.05 &    0.10 \\
\ion{Mn}{1} &    0.09 & $-$0.02 & $-$0.01 &    0.15 &    0.18 \\
\ion{Fe}{1} &    0.08 & $-$0.01 & $-$0.12 &    0.09 &    0.17 \\
\ion{Fe}{2} &    0.01 &    0.06 & $-$0.02 &    0.05 &    0.08 \\
\ion{Co}{1} &    0.09 & $-$0.01 & $-$0.01 &    0.07 &    0.11 \\
\ion{Ni}{1} &    0.10 & $-$0.04 & $-$0.19 &    0.09 &    0.24 \\
\ion{Sr}{1} &    0.06 &    0.05 & $-$0.08 &    0.10 &    0.15 \\
\ion{Ba}{1} &    0.07 &    0.08 & $-$0.05 &    0.10 &    0.15 \\
\enddata
\tablenotetext{a}{Calculated from the quadratic sum of the individual error estimates.}
\end{deluxetable*}

\end{document}